\documentclass[preprintnumbers,amsmath,amssymb,aps,prd,notitlepage,11pt,tightenlines]{revtex4-1}
\pdfoutput=1

\usepackage{amsfonts,amssymb,amsmath,mathbbol,dsfont}
\usepackage{graphicx}
\usepackage{multirow}
\usepackage{subfig,floatrow}
\usepackage{feynmp}
\usepackage{color}
\usepackage[pdftex]{hyperref}
\newcommand\one{\leavevmode\hbox{\small1\normalsize\kern-.33em1}}

\newcommand{\lag}{\mathcal{L}}
\newcommand{\ord}{\mathcal{O}}
\newcommand{\ope}{\mathcal{O}}

\newcommand{\qqqquad}{\qquad \qquad \qquad}

\newcommand{\gev}{{\ensuremath\rm GeV}}
\newcommand{\tev}{{\ensuremath\rm TeV}}

\newcommand{\ifb}{{\ensuremath\rm fb^{-1}}}

\def\slashchar#1{\setbox0=\hbox{$#1$}           
   \dimen0=\wd0                                 
   \setbox1=\hbox{/} \dimen1=\wd1               
   \ifdim\dimen0>\dimen1                        
      \rlap{\hbox to \dimen0{\hfil/\hfil}}      
      #1                                        
   \else                                        
      \rlap{\hbox to \dimen1{\hfil$#1$\hfil}}   
      /                                         
   \fi}

\def\eg{{\sl e.g.} \,}

\newcommand{\be}{\begin{eqnarray*}}
\newcommand{\ee}{\end{eqnarray*}}

\newcommand{\bee}{\begin{eqnarray}}
\newcommand{\eee}{\end{eqnarray}}
\newcommand{\beeq}{\begin{equation}}
\newcommand{\eeeq}{\end{equation}}

\begin{document}

\begin{fmffile}{Feynman}

\title{Polarized WW Scattering on the Higgs Pole}

\author{Johann Brehmer}
\author{Joerg Jaeckel}
\author{Tilman Plehn}
\affiliation{Institut f\"ur Theoretische Physik, Universit\"at Heidelberg}

\begin{abstract}
  The Higgs discovery has given us the Higgs--gauge sector as
  a new handle to search for physics beyond the Standard Model. This 
  includes physics scenarios originally linked to 
  massive gauge boson scattering at high energies. We
  investigate how one can separately probe the Higgs couplings to the
  longitudinal and transverse parts of the massive gauge bosons away
  from this high-energy limit. Deviations from the Standard Model could
  originate from higher-dimensional operators,
  compositeness, or even more fundamentally from a violation of gauge
  invariance.  The signature we propose is the tagging jet kinematics
  in weak boson scattering for scattering energies close to the Higgs
  resonance. During the upcoming LHC run at 13~TeV we will be able to
  test these couplings at the 20\% level.
\end{abstract}

\preprint{1404.5951}

\maketitle

\tableofcontents

\section{Introduction}
\label{sec:intro}

With the discovery of a scalar Higgs boson~\cite{higgs,discovery} the
Standard Model of particle physics is finally complete. However, many
experimental and theoretical questions are left
un-answered~\cite{bsm_review}: not least among them is the question
about the smallness of the Higgs mass, or more precisely the
electroweak scale, when compared to the Planck scale. Are the
interactions of the most fundamental particles really described by
perturbative field theories all the way to the Planck
scale~\cite{sm_only}? Also the symmetry structure of the SM itself is
surprising; why do we observe an SU(3)$\times$SU(2)$\times$U(1) gauge
symmetry?

One way to make progress on these questions is theoretical model
building. At the same time, however, it is crucial that we check the
underlying assumptions of the Standard Model by experimentally testing
the predictions following from its basic structure. In that one should
leave no stone unturned, including tests of the nature of gauge
invariance itself.\bigskip

The discovery of the Higgs opens exciting new possibilities to test
the Standard Model, in particular with the above questions in
mind. The scalar nature of the Higgs is responsible for the hierarchy
problem, and indeed the Higgs is the only scalar which could be a
fundamental non-composite particle. At the same time the
Higgs mechanism is responsible for generating the masses
of the $W$ and $Z$ bosons and is therefore intimately linked to the
SU(2)$\times$U(1) gauge structure. The Higgs--gauge sector therefore provides
an ideal laboratory to probe both the hierarchy problem and 
the gauge structure of the Standard Model.  Specifically,
measurements of Higgs couplings to vector bosons as probed in $VV$
scattering processes ($V=W,Z$) are ideal for this endeavor.

Historically, the question used to be how $VV$
scattering is unitarized or how its amplitude behaves at
high energies. The corresponding experimental signature is 
$VV$ scattering at high energies, described by the Goldstone 
equivalence theorem. By now we know that the observed Higgs boson
will at least significantly dampen the ultraviolet behavior of $VV$
scattering~\cite{unitarity}. The questions 
\begin{enumerate}
\item How does the $VV$ scattering amplitude behave at high energies?
\item Does the observed Higgs boson render $VV$ scattering weakly interacting?
\item Does the $HVV$ interaction correspond to the Standard Model prediction?
\end{enumerate}
are equivalent in their theoretical implications, even though they
seem to require very different experimental strategies.  In this paper
we propose to follow the third question, because there exists plenty
of experimental data from Higgs analyses which allows us to test the
structure of the $HVV$ vertex to gain insight into the underlying
fundamental physics.\bigskip

The Higgs--gauge sector is often approached using an effective field
theory approach in terms of higher-dimensional operators. While it
follows naturally from the observation that gauge field theories
describe the fundamental interactions of particles~\cite{d6_review},
the greatest advantage of this approach is its greatest shortcoming:
it usually assumes (part of) the symmetry structure of the Standard
Model and hence makes it harder to study the fundamental symmetries
underlying the Standard Model. 

In this paper we are a bit more heretical: we propose to
\emph{separately} test the Higgs couplings to transverse and
longitudinal parts of the massive gauge bosons. This is inspired by
the approximate notion that the transverse parts correspond to the
`proper' gauge bosons, whereas the longitudinal parts arise from the
eaten Goldstone bosons.  
A clean separation is only possible at
infinitely large momenta or in the nearly massless
limit~\cite{equivalence}, at which we cannot operate when interpreting
LHC Higgs data. However, this technical problem does not render the 
physics question of coupling massive gauge bosons with different 
polarizations to the Higgs boson any less relevant.
Indeed, through the explicit definition of polarization in a simple model we
have to break Lorentz invariance.
Nevertheless, we view it as a first step
towards individually testing the physics of the Higgs and its
Goldstone bosons and that of the original non-Higgsed gauge bosons.

On the theory side, composite Higgs bosons are studied in great numbers
and detail. However, much less attention has been paid to the known
possibility that the gauge bosons can be composite particles of an
emergent gauge
symmetry~\cite{gauge_composite}. As an
example, vector meson dominance in QCD can be viewed as an
(approximate) emergent gauge
symmetry~\cite{Komargodski:2010mc}. Moreover, supersymmetric theories
often have completely different gauge symmetries in the infrared and
in the ultraviolet, as made explicit by the Seiberg
duality~\cite{Seiberg:1994pq}. The gauge symmetry can be larger in the
infrared than in the ultraviolet regime and therefore does not simply
correspond to a Higgsed case. Indeed, the new degrees of freedom can
be viewed as solitons of the original theory and are non-trivial in
this sense.  In this paper we take the first steps to develop tools to
experimentally distinguish between a composite Higgs sector and
composite gauge bosons. The question to how well our question and 
our findings can be described in terms of higher-dimensional operators
will be discussed at the end of the paper.\bigskip

\begin{figure}[t]
\begin{center}
\includegraphics[width=0.5 \textwidth]{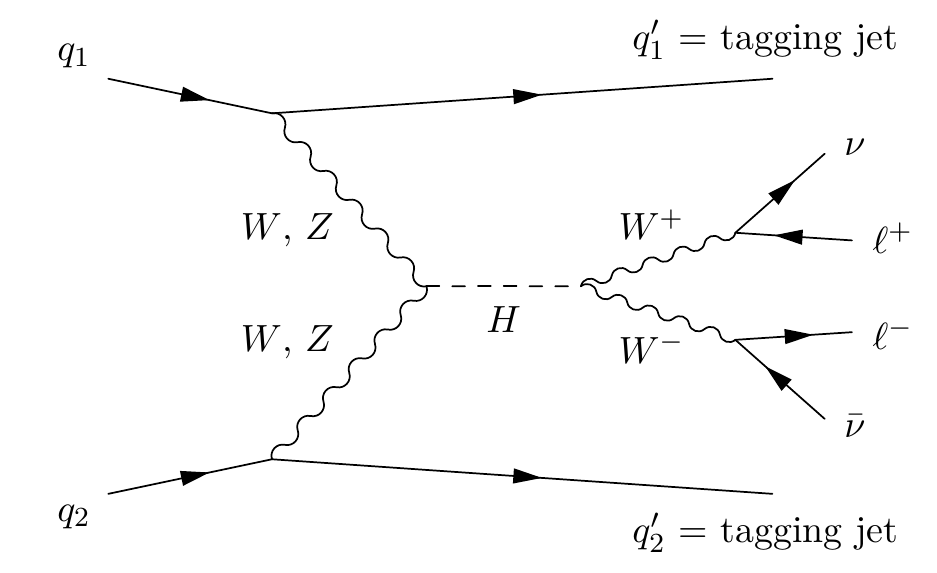}
\end{center}
\caption{$VV$ scattering process we consider in this paper, $V=W$. Our
  analysis shows that indeed the Higgs resonance region is where we
  can best test the Higgs--gauge structure in and beyond the Standard Model.}
  \label{fig:VVscattering}
\end{figure}

Our laboratory to probe the structure of the $HVV$ interaction is
$VV$ scattering as depicted in Figure~\ref{fig:VVscattering}.
At large energies this process has been
discussed extensively~\cite{Cheung:2008zh}.  For example in the
Refs.~\cite{Bagger} the authors
develop analysis strategies to select highly energetic longitudinal
gauge bosons in all leptonic $VV$ topologies, including jet tagging, a
central jet veto, and back-to-back lepton geometries.
In Ref.~\cite{Butterworth} semileptonic $W^+W^-$ decays are first
analyzed based on boosted $W$ tagging.  The full set of semileptonic
$VV$ channels is discussed in Ref.~\cite{Ballestrero}. As pointed out
in Ref.~\cite{tao} these high-energy $VV$ signatures suffer not only
from low rates, but also from large scale uncertainties. The
measurement of the relative longitudinal and transverse polarizations
based on decay angles can be used to avoid large QCD uncertainties on
LHC rates.

In contrast to all this earlier work our main focus lies on
weak--boson--fusion production of massive gauge bosons close to the
Higgs resonance, which enhances the event rates. Instead of using the
high-energy behavior of the vector bosons, we look at the tagging jets
and in particular their high-energy behavior.
Beyond the interpretation in specific models our analysis therefore
develops a set of new observables that
are sensitive to the structure of the Higgs--gauge sector.\bigskip

The paper is structured as follows: in Section~\ref{sec:model} we
start with our definition of transverse and longitudinal gauge
bosons. Then, in Section~\ref{sec:wbf} we develop our analysis
strategy and contrast it with the established studies of the
high-energy regime and the Higgs decay correlations. The results of our
tagging jet analysis follow in Section~\ref{sec:results}. 
Section~\ref{sec:frame} is devoted to
a discussion of the frame dependence of polarizations.
Finally, in Section~\ref{sec:dim6} we relate our findings to the usual effective
field theory approach based on higher-dimensional Higgs--gauge
operators.

\section{A simple model}
\label{sec:model}

Even in the absence of an immediate field theoretical description, the
polarization of massive spin-1 gauge bosons coupling to the Higgs
bosons is a property of the $W$ and $Z$ bosons worth testing at the
LHC. As alluded to in the introduction, the coupling of the Higgs
boson to transverse and longitudinal gauge bosons has a very different
origin. This aspect is obvious when we go into the high-energy limit,
where the longitudinal degrees of freedom correspond to the Goldstone
modes of a Higgs sigma field and where the transverse polarization
modes are suppressed by powers of $M_V/E$. At finite energies and for
example in unitary gauge we have to define and separate the
transverse and longitudinal degrees of freedom by hand.

In spite of this complication, measuring the polarization of the gauge
bosons coupling to the Higgs will allow us to test the underlying
structure of electroweak symmetry breaking. The $HVV$ couplings
($V=W,Z)$ mediating both, Higgs production in weak boson fusion and
Higgs decays to leptons, require us to define the polarization of the
gauge bosons for off-shell and on-shell states. We define the
transverse and longitudinal parts of the $W$- or $Z$-boson fields as
\begin{equation}
V_T^\mu = \mathbb{P}_\nu^\mu \; V^\nu
\qquad \text{and} \qquad
V_L^\mu = \left( \mathds{1} - \mathbb{P} \right)^\mu_{\,\nu} \; V^\nu \; ,
  \label{eq:TransverseDefinition}
\end{equation}
in terms of the $W,Z$ boson in unitary gauge and the projection
operator $\mathbb{P}^{\mu\nu}$
\begin{alignat}{5}
\label{projectors}
\mathbb{P}^{0\nu} = 0 = \mathbb{P}^{\mu0}  
\qquad \text{and} \qquad 
\mathbb{P}^{ij} = \delta^{ij} - \frac {\vec{p}^i \vec{p}^j} {\vec{p}^2} 
\qquad (i,j=1,2,3) \; .
\end{alignat}
This allows us to split the Higgs--gauge vertex into its polarization
components. Measuring polarizations requires a specific reference
frame. We choose to evaluate Eq.~\eqref{eq:TransverseDefinition} in the
Higgs rest frame. This definition, which we will justify below, gets rid of
mixed couplings $H V_LV_T$ in
the Higgs--gauge coupling structure $H V_L V_L + H V_T V_T$. The
remaining two contributions to the $HVV$ coupling can be written in
terms of the Standard--Model coupling strength $g_\text{SM}$ and two
scaling parameters
\begin{equation}
\lag \supset - g_\text{SM} \; H \; 
\left( a_L \;  V_{L\,\mu} V_L^\mu + a_T \; V_{T\,\mu} V_T^\mu 
\right) \; .
\label{eq:lag_pol}
\end{equation}
The sign of the real scaling parameters $a_L$ and $a_T$ is free, and
we do not enforce a sum rule to protect the total Higgs production and
decay rates.\bigskip

As already mentioned, this simple model in terms of transverse and
longitudinal polarizations requires a choice of reference frame and
therefore breaks Lorentz invariance. Our simple model is also clearly not
gauge-invariant. However, independent longitudinal and transverse
Higgs--gauge couplings can be induced by perfectly valid models respecting
the symmetries of the Standard Model. We will demonstrate this
in Section~\ref{sec:dim6}, where we link our simple model to higher-dimensional
operators. In this effective field theory approach, the couplings $a_L$ and $a_T$
become momentum-dependent. The correspondence to our simple model
is most obvious if we define polarizations in the Higgs rest frame,
justifying the choice made above. Alternative definitions will be
discussed in Section~\ref{sec:frame}.

Regardless of these complications, the simple model exhibits a
straightforward correspondence to the appropriate definition of
longitudinal and transverse gauge bosons in the high-energy limit,
which is precisely what we want to probe.  We will therefore use it as
a test scenario for our LHC analysis.

\section{Tagging jet kinematics}
\label{sec:wbf}

Higgs production in weak boson fusion probes the $HVV$ coupling
structure in the initial state. Similarly, the $VV$ decays test the
same coupling in the final state. Due to the dependence on the
initial state couplings the energy scales probed by
the initial vertex are not automatically limited to the Higgs
mass, as is the case for Higgs decays. Without any requirements on the
intermediate Higgs state, the relevant process is $W$ pair production
in weak boson fusion at the perturbative order $\ord(\alpha^4)$,
\begin{equation}
  p p \to W^+W^- \, jj \to (\ell^+ \bar \nu) \, (\ell^- \nu) \, jj \; .
  \label{eq:process}
\end{equation}
The on--shell Higgs diagram contributing to this process is shown in
Figure~\ref{fig:VVscattering}.  Because the observed Higgs boson has
essentially Standard--Model--strength couplings to the weak gauge
bosons, we know that a large fraction of the rate for the full process
in Eq.~\eqref{eq:process} comes through the $s$-channel Higgs
resonance. In that case one of the two $W$ bosons will be far off its
mass shell and the Higgs resonance can be extracted with a transverse
mass variable. This channel is maximally sensitive to the $HWW$
coupling structure, which governs both the Higgs production and decay.

We generate event samples for the process given in
Eq.~\eqref{eq:process} assuming an LHC energy of 13~TeV with
\textsc{MadGraph5}~\cite{mg5} using an in-house implementation of the
model described in Section~\ref{sec:model}.
Because it is clear that the weak--boson--fusion features
can be experimentally extracted~\cite{atlas_wbf}, we limit ourselves
to the parton level and omit systematic uncertainties in this first study.
For the Higgs signal we assume a
Higgs mass of 125~GeV and a Higgs width of 4.4~MeV, calculated with
\textsc{HDecay}~\cite{hdecay}. The polarizations of the $W$ and $Z$
bosons are defined according to
Eq.~\eqref{eq:TransverseDefinition}.\bigskip

\begin{figure}[t]
\includegraphics[width=0.45 \textwidth]{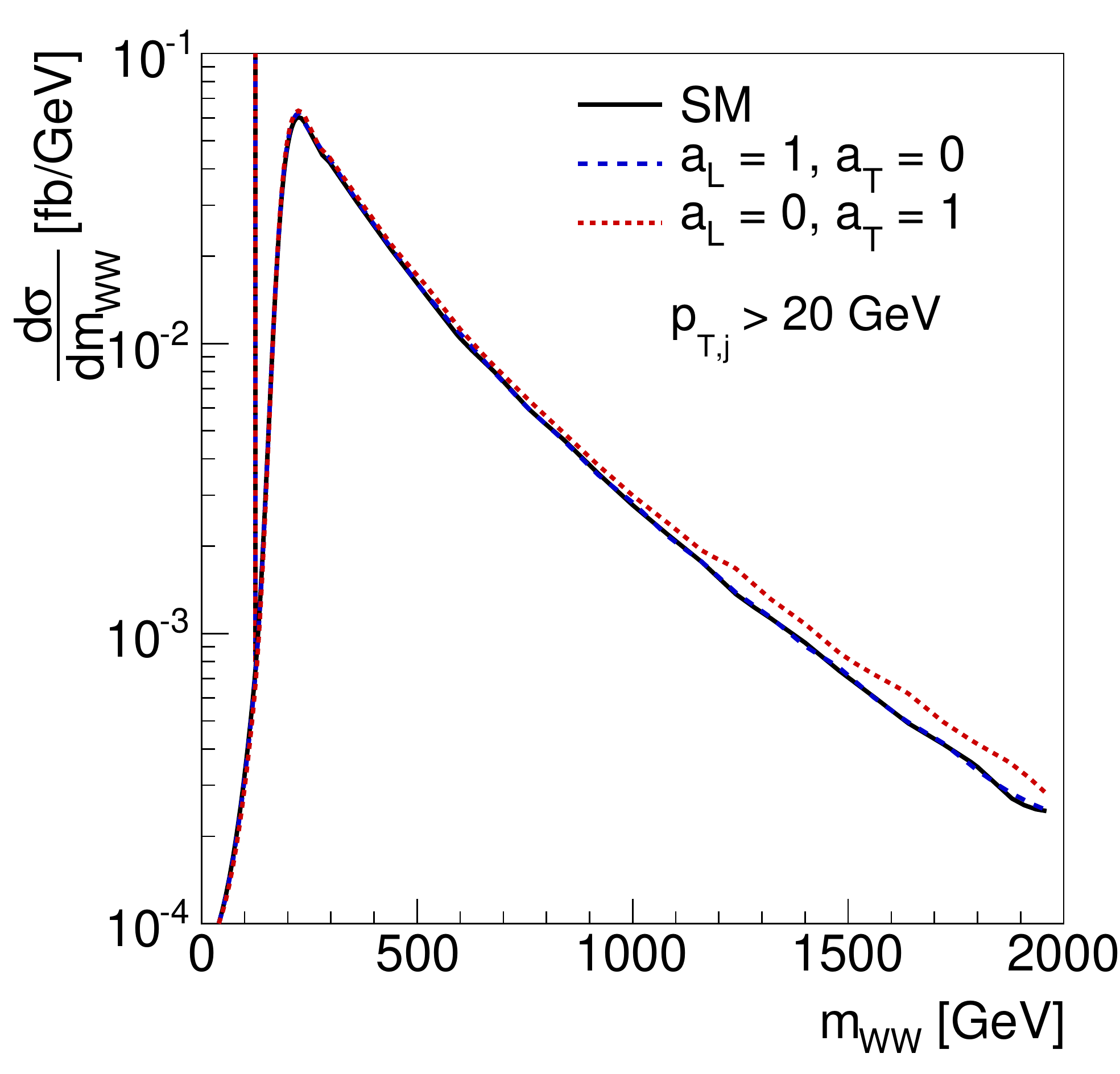}
\hspace*{0.05\textwidth}
\includegraphics[width=0.45 \textwidth]{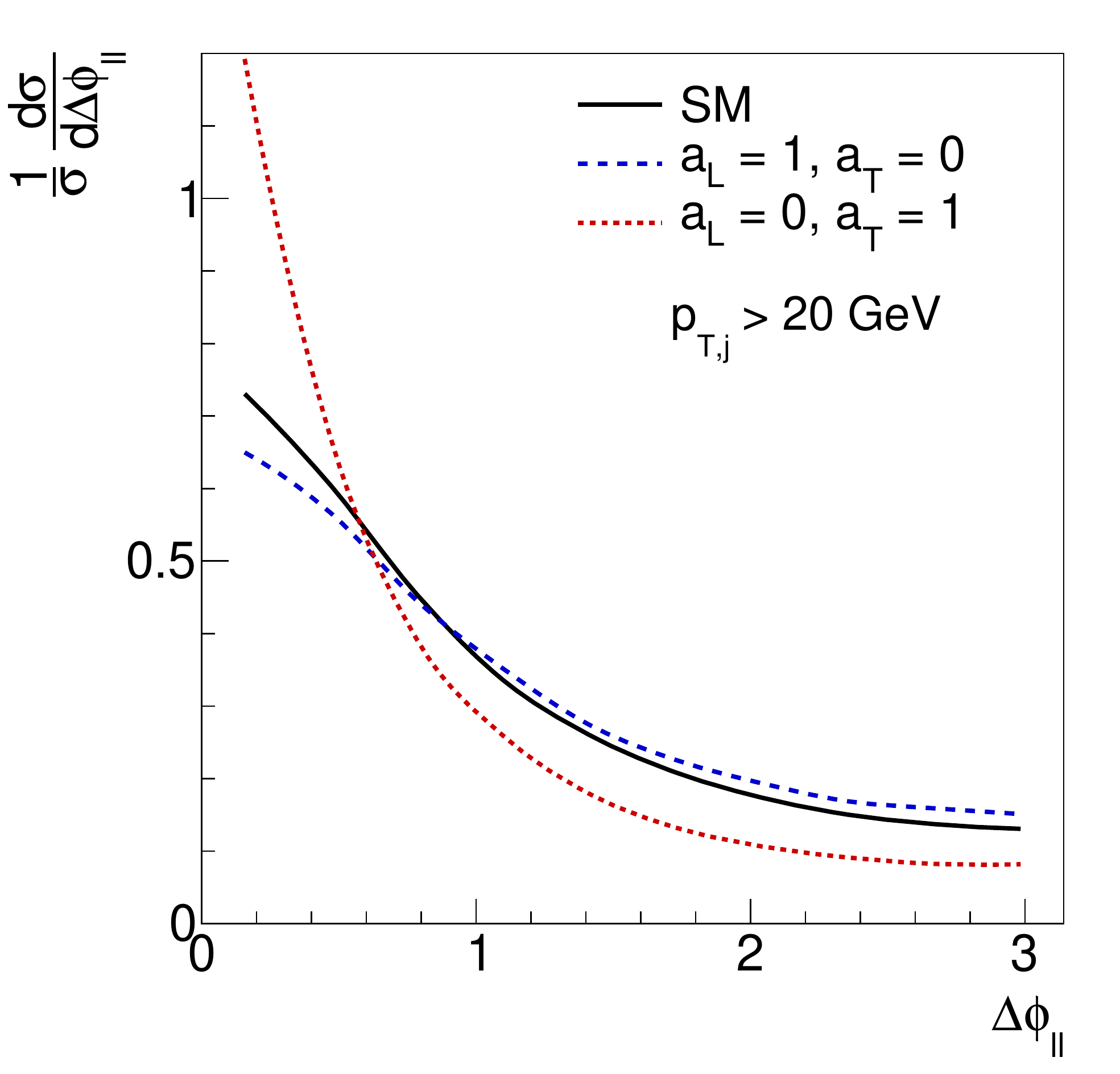}
\caption{Left: $m_{WW}$ distribution for the signal process defined in
  Eq.~\eqref{eq:process} for a Higgs coupling exclusively to transverse
  or longitudinal gauge bosons as well as the Standard Model
  case. Right: $\Delta \phi_{\ell\ell}$ distributions of the Higgs
  decay leptons for the same three setups. We require $p_{T,j} >
  20$~GeV to remove photon exchange contributions.}
  \label{fig:alternative}
\end{figure}

Before we start with our tagging jet analysis we need to briefly
motivate two choices which we make in analyzing the signal process
given in Eq.~\eqref{eq:process}, namely that we
\begin{enumerate}
\item focus on the Higgs pole rather than on the high-energy regime, and
\item analyze the tagging jets in addition to decay correlations.
\end{enumerate}
The analysis of the process given in Eq.~\eqref{eq:process} in the
search for non-standard strong interactions in the gauge sector has a
long tradition~\cite{Bagger,Butterworth}. The main observable in such analyses
is the invariant mass of the $WW$ system or any approximation to
it. In the high-energy limit this process should follow the
equivalence theorem~\cite{equivalence} and be dominated by the
Goldstone modes of the massive $W$~bosons. This is exactly the
behavior we see in the left panel of Figure~\ref{fig:alternative},
where we simulate events for weak--boson--fusion $W$ pair production.
In the setup where the Higgs couples only to transverse gauge bosons
we observe a significant rate enhancement for $m_{WW} \gtrsim
1$~TeV. This is due to the missing cancellation with the scattering of
longitudinal gauge bosons mediated by a Higgs exchange. In contrast, a
Higgs coupling only to longitudinal gauge bosons is indistinguishable
from the Standard Model case, showing that little cancellation is at
work. In the same distribution we can also see the main problem with
such an analysis: the fraction of events populating the phase space
regions sensitive to modified Higgs--gauge couplings is depressingly
small.

Moreover, after the Higgs discovery the motivation for this kind of
challenging analysis is less straightforward. First, we can assume that the
ultraviolet behavior of $WW$ scattering is at least to a large part
cured by the observed Higgs boson. This postpones any non-trivial behavior 
to much higher energies and therefore lower cross sections. Second, we 
can measure the Higgs
properties, like the structure of the $HVV$ vertex and its coupling
strength, and based on those results reliably predict the behavior of
the $WW$ scattering at high energies. Given that most of the rate for
$qqWW$ production comes from the Higgs pole region it is natural to
replace the analysis of strongly interacting $WW$ scattering by a
dedicated study of $qqWW$ production around the Higgs pole. An explicit
study of the actual $WW$ scattering process is of course still welcome
and interesting, but would require a different interpretation.\bigskip

The second choice is less obvious.  Clearly, the Higgs decay leptons
and the missing transverse momentum provide a handle to the final $WW$
system. As a matter of fact, the LHC analyses leading to the Higgs
discovery~\cite{discovery} already use a correlation between the two
leptons reflecting the scalar nature of the intermediate
resonance~\cite{lepton_angle}.  The most straightforward observables
related to polarizations are decay angles~\cite{tao}. In the presence
of two neutrinos alternative observables include the lepton transverse
momenta, the dilepton mass $m_{\ell \ell}$, the azimuthal angle
between the leptons $\Delta \phi_{\ell \ell}$, and the missing
transverse momentum $p_T^\text{miss}$. For illustration purposes we
show the $\Delta \phi_{\ell\ell}$ distribution
for the process $qq \to qq H \to qq (W^+W^-)$
in the right panel of Figure~\ref{fig:alternative}. Unlike for the
$m_{WW}$ distribution the distinguishing features now reside in a
phase space region with a significant number of events. The main
question in this case becomes how much of this difference survives
acceptance cuts, background rejection cuts, and detector
effects~\cite{tao}.

Instead of the Higgs decay products we will focus on the tagging jets.
Their kinematic distributions are sensitive to the initial $HVV$ vertex
structure~\cite{delta_phi_orig,delta_phi} and the polarization of the
fusing gauge bosons. While originally developed for weak boson fusion,
the same features can be used for example in Higgs
production~\cite{higgs_spin,delta_phi_gg} or in top pair production in gluon
fusion~\cite{delta_phi_pairs}.
In the following we will see that this tagging jet information
includes the same kind of information on the gauge boson polarization
as the Higgs decay correlation. However because it is independent of
the Higgs decay channel, it can be measured in many different
signatures, allowing for an efficient test of the 
results from individual Higgs decay channels.\bigskip

We can analytically formulate the relation between
polarization of the weak bosons and the transverse momenta of the
tagging jets in the effective $W$ approximation
(EWA)~\cite{ewa_orig}. It treats the initial gauge bosons as on-shell
propagators and factorizes the weak--boson--fusion process into the
radiation of a collinear $W$ off a quark and the hard scattering
process of the two $W$~bosons.  Similar to a $p_T$-dependent
leading-order QCD parton density we can define a probability of
finding a $W$~boson radiated off the incoming quark with a
longitudinal momentum fraction $x$ and the transverse momentum
$p_T$. For the respective polarizations they read
\begin{alignat}{5}
P_T(x, p_T) &= \frac{g^2}{16 \pi^2} \; \frac{1 + (1-x)^2}{x} \; \frac{p_T^3}{((1-x)m_W^2 + p_T^2)^2} \notag \\ 
P_L(x, p_T) &= \frac{g^2}{16 \pi^2} \; \frac{1-x}{x} \; \frac{2 (1-x) m_W^2 p_T}{((1-x)m_W^2 + p_T^2)^2} \; .
\label{eq:ewa}
\end{alignat}
The two distributions have a different $p_T$ dependency: tagging jets
recoiling against a longitudinal $W$ should be softer than those
recoiling against a transverse $W$.\bigskip

\begin{figure}[t]
\includegraphics[width=0.45 \textwidth]{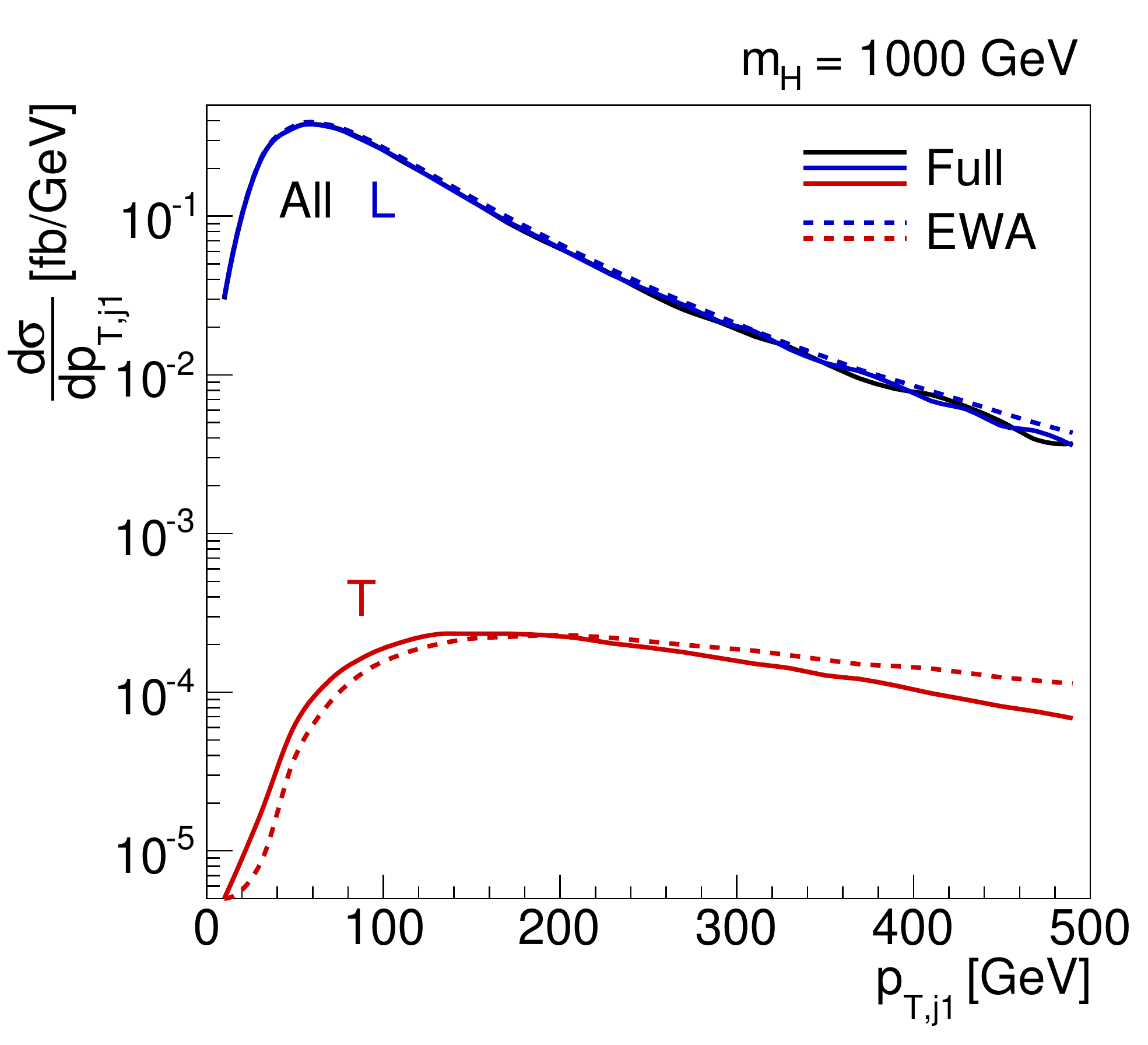}
\hspace*{0.05\textwidth}
\includegraphics[width=0.45 \textwidth]{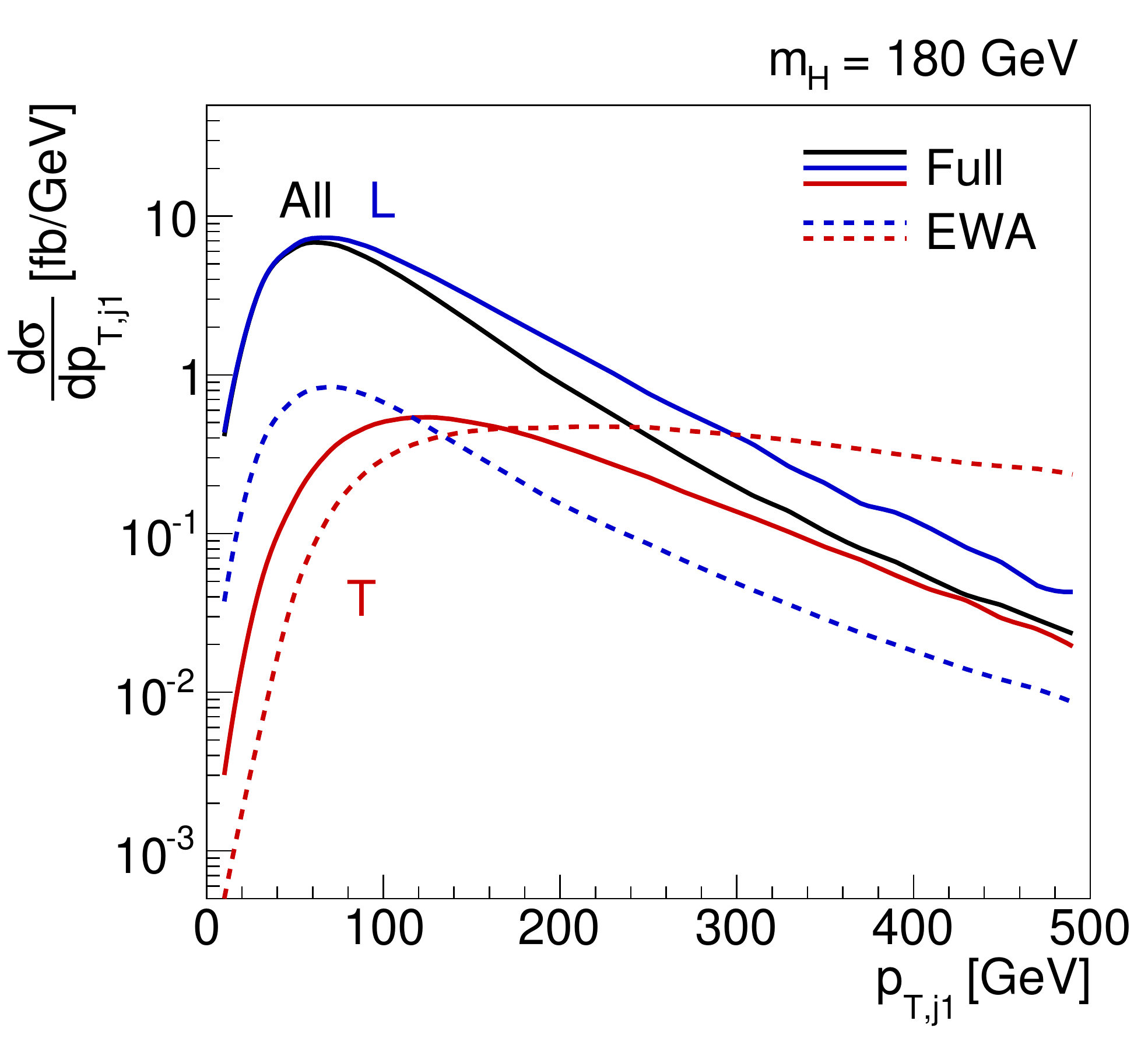}
\caption{Normalized $p_{T,j1}$ distributions for the complete signal
  process defined in Eq.~\eqref{eq:process} and in the effective $W$
  approximation, assuming a heavy Higgs with $m_H = 1$~TeV (left) and
  $m_H=180$~GeV (right).}
  \label{fig:ewa}
\end{figure}

The theoretical predictions shown in Eq.~\eqref{eq:ewa} are a promising
basis to test the longitudinal versus transverse structure of the $HVV$
coupling in weak boson fusion. However, we first need to see if the
assumptions underlying the effective $W$ approximation actually
hold. To consider the $W$ to be essentially a parton inside the proton there
needs to be a hierarchy $E_\text{proton} \gg m_\text{hard} \gg m_W$.
In our case the hard process is the $2\to 2$ scattering $W^+W^- \to
W^+W^-$, where the two relevant Mandelstam variables scale like $|t| <
s = m_\text{hard}^2$. For the full $2 \to 4$ process the additional
two energy scales defined by the tagging jet momenta then have to be
$p_{L,j}\gg m_\text{hard} \gg p_{T,j} \sim
m_W$~\cite{tagging,wbf_tau,wbf_w}.  

The physical Higgs mass of $m_H=125$~GeV is too close to the $W$ mass
and the conditions for the validity of the EWA are not fulfilled.  We
nevertheless find that while the EWA is not valid quantitatively, 
the qualitative behavior is similar.

To investigate the validity of the EWA we compare the EWA to the full 
result for two different assumed Higgs masses above the $W$ pair 
production threshold.
In the left panel of
Figure~\ref{fig:ewa} we show the $p_T$ distributions for the two
tagging jets for an assumed Higgs mass of 1~TeV. Only the signal
subprocess $ud \to du H$ on the Higgs mass pole is included.  Indeed,
the distributions for both, transverse and longitudinal $W$ and $Z$
polarizations agree well with the EWA predictions in
Eq.~\eqref{eq:ewa}. Note that for the transverse part even a Higgs mass
of 1~TeV shows worse agreement for higher momenta, suggesting a
different fall--off behavior.  For lighter Higgs bosons the agreement
becomes gradually worse, as can be seen for $m_H=180$~GeV in the right
panel of Figure~\ref{fig:ewa}. Below the threshold value $m_H = 2 m_W$
the EWA assuming incoming on-shell $W$~bosons loses its
validity. Nevertheless, the
effective $W$ approximation motivates a study of the transverse
momenta of the tagging jets to extract information on the polarization
of the initial $W$ and $Z$ bosons.\bigskip

For our actual analysis we use the full amplitude given by
Eq.~\eqref{eq:process}. As a first background this includes off-shell
$W$ pair production at the same purely electroweak order in
perturbation theory. One irreducible background is given by the same
initial and final state and with an intermediate Higgs boson, but
coupling the Higgs to two gluons at order $\ord(\alpha_{ggH}
\alpha_s^2 \alpha)$.  An additional non-Higgs background is $W$ pair
production in association with two jets at order
$\ord(\alpha_s^2\alpha^2)$~\cite{wbf_w}. The kinematics of the two
tagging jets~\cite{tagging} and the structure of additional central
QCD radiation~\cite{cjv} can be used to suppress these
backgrounds. In particular through the additional jet activity
there exist many ways to reduce the $t\bar{t}+{}$jets
background~\cite{wbf_w}  and we therefore omit it in this first study and limit
ourselves to the more dangerous $W^+ W^-$ processes on and off the
Higgs mass shell. We start by requiring that the $W$ decay leptons
($\ell = e, \mu$) satisfy the staggered cuts
\begin{alignat}{7}
| \eta_\ell | < 2.5 
\qqqquad 
p_{T,\ell} > 20,10~\gev
\qqqquad 
p_T^\text{miss} > 20~\gev \; .
\label{eq:cuts_lep}
\end{alignat}
The two forward partons forming the tagging jets have to fulfill the
standard weak--boson--fusion cuts~\cite{wbf_tau,wbf_w}
\begin{alignat}{5}
| \eta_j | &< 5.0 
& p_{T,j} &> 25~\gev \notag \\
\Delta \eta_{jj} &> 4.2
\qqqquad 
& m_{jj} &> 500~\gev
\qqqquad 
& \eta_{j_1} \cdot \eta_{j_2} &< 0 \; .
\label{eq:cuts_jets}
\end{alignat}
Following the definition of longitudinal and transverse $W$ and
$Z$ bosons in Eq.~\eqref{eq:lag_pol} we simulate 529 parameter points
in the $(a_L, a_T)$ space. Of these 441 are evenly distributed in the range
$a_{L/T} \in [-2,2]$, while 88 increase the
sensitivity close to the Standard--Model value $a_L = a_T = 1$. For
each parameter point we generate approximately $10^5$ events. In
between these points we interpolate cross sections and $p$-values by
Delauny triangulation.\bigskip

\begin{figure}
  \begin{floatrow}
    \floatbox{figure}[0.55\textwidth][\FBheight][t] 
    {\caption{Transverse mass distribution for the signal and backgrounds
        in the Standard Model. The dashed lines indicate our selection
        cuts.}
      \label{fig:mt}}
    {\includegraphics[width=0.60\textwidth]{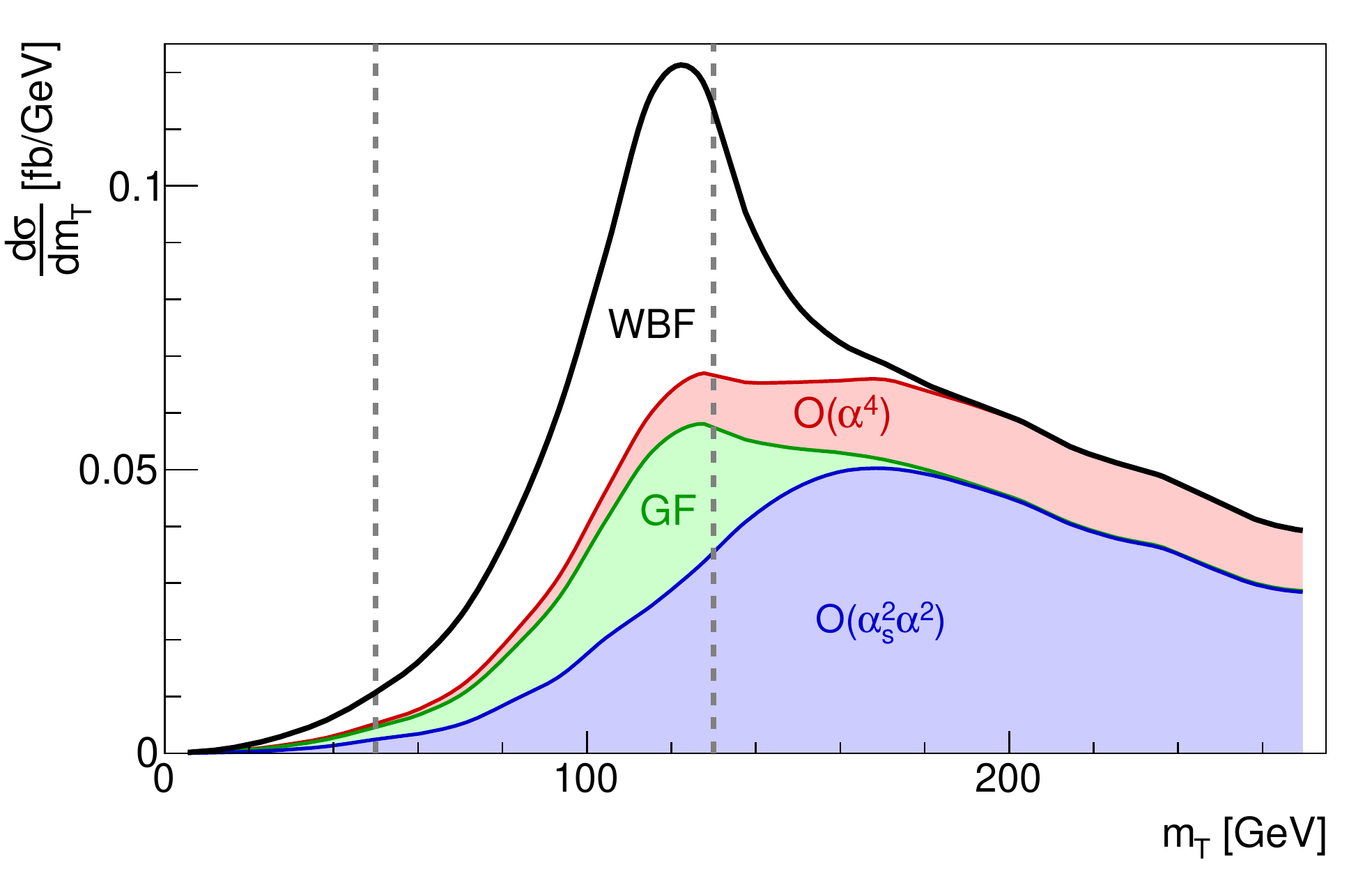}}
    \hspace*{0.4cm} \floatbox{table}[0.4\textwidth][\FBheight][t]
    {\caption{Corresponding cross sections for the different
        sub-processes before and after the $m_T$ cut.}
        \label{tab:mt}}
      {\begin{footnotesize}
            \vspace*{0.35cm}
            \begin{ruledtabular}
              \begin{tabular}{lrr}
              \ & all & $m_T$ cut \\
              \hline
              WBF $H \to W^+W^-$ & 3.15 & 2.34 \\
              Continuum $\ord(\alpha^4)$ & 4.54 & 0.31 \\
              GF $H \to W^+W^-$ & 1.62 & 1.13 \\
              Continuum $\ord(\alpha_s^2 \alpha^2)$ & 11.01 & 1.17 \\
              \hline
              $S/(S + B)$ & 0.15 & 0.47
            \end{tabular}
            \end{ruledtabular}
        \end{footnotesize}
      }
    \end{floatrow}
  \end{figure}

To select events from the Higgs resonance, we use a transverse mass
variable $m_T$, which we define using the approximation $m_{\nu \nu}
\approx m_{\ell \ell}$~\cite{wbf_w,lecture}.
In Figure~\ref{fig:mt} we show
the different contributions to the transverse mass distribution. The
selection cut
\begin{equation}
  50~\gev <  m_{T,WW} < 130~\gev
  \label{eq:mTCut}
\end{equation}
retains $74\%$ of the Higgs resonance events, leading to a
signal--to--background ratio around unity. The Standard Model
cross sections before and after this cut for the different processes
are given in Table~\ref{tab:mt}. For the remainder of this work, we
require the transverse on-shell condition of Eq.~\eqref{eq:mTCut} in
addition to the lepton cuts of Eq.~\eqref{eq:cuts_lep} and tagging jet
cuts of Eq.~\eqref{eq:cuts_jets}.

\section{Results for the simple model}
\label{sec:results}

Following the discussion in the last section we will focus on Higgs
production in weak boson fusion, with a subsequent decay to a leptonic
$W^+W^-$ pair.  Based on Eq.~\eqref{eq:lag_pol} we define Higgs
couplings to transverse and longitudinal $W$ and $Z$ bosons, both
contributing to the Higgs signature. In the following we will answer
the question which observables will allow us to constrain the two
coupling parameters $a_L$ and $a_T$, both normalized to the Standard
Model value $a_{L,T} =1$, at the upcoming LHC run.\bigskip

\begin{figure}[t]
\centering
\includegraphics[width= 0.43 \textwidth]{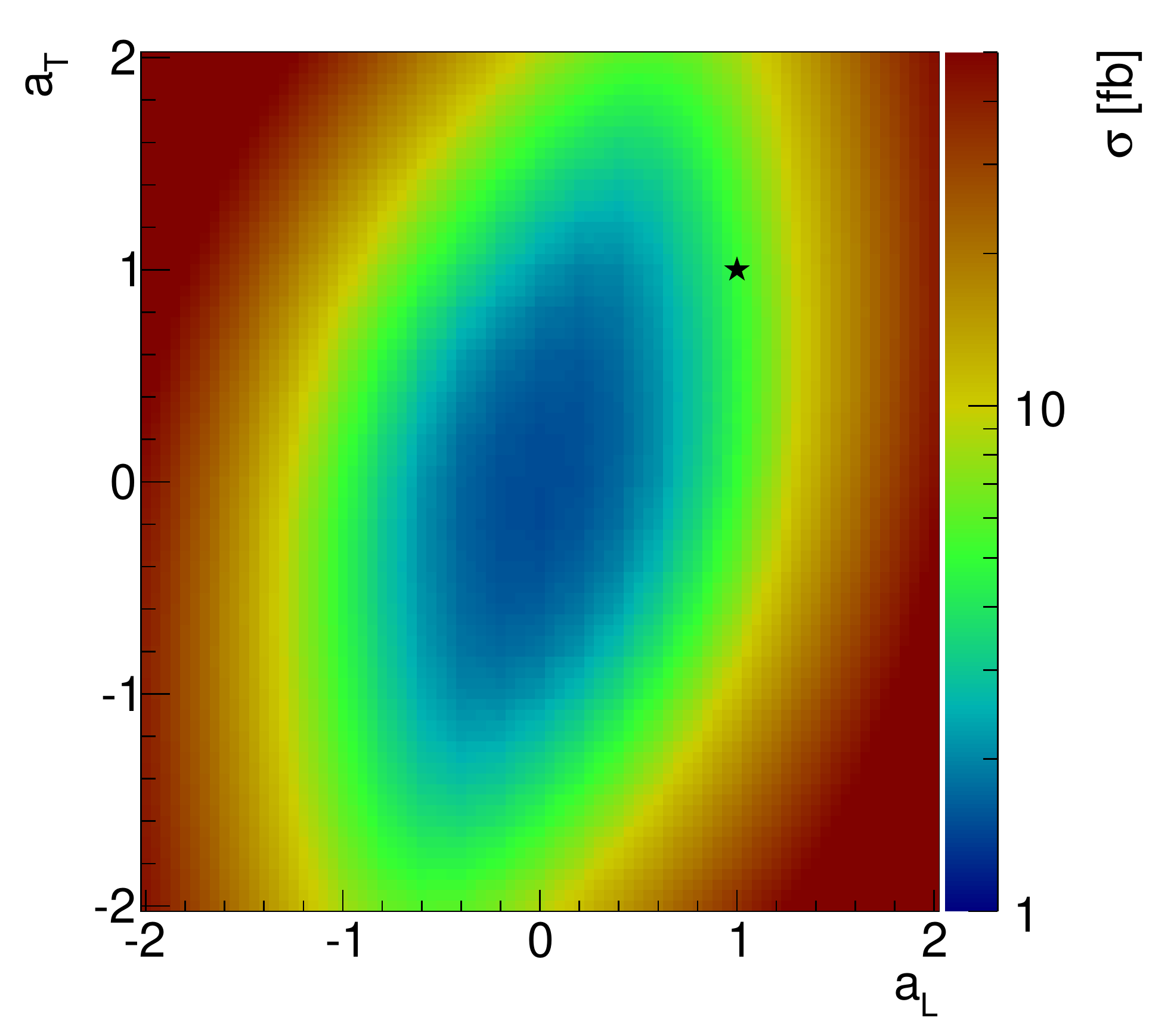}
\caption{Cross sections on the Higgs resonance for various couplings
  of the Higgs boson to longitudinal and transverse vector bosons. The
  Standard Model is marked with a star.}
\label{fig:XSec}
\end{figure}

\paragraph{Total rate:}
The first observable we analyze is the total cross section on the
Higgs resonance, measured as the signal strength in the two--jet
category both by ATLAS and by CMS~\cite{ex_ww}.  In
Figure~\ref{fig:XSec} we give this cross section after cuts,
Eqs.~\eqref{eq:cuts_lep}\;--\;\eqref{eq:mTCut}, as a function of $a_L$ and
$a_T$.
The curve with a constant cross section is approximately
an ellipse in the $(a_L,a_T)$ plane.
Close to the Standard Model the rate is nearly insensitive to the transverse
coupling. This reflects the fact that the Higgs boson in the Standard
Model couples predominantly to longitudinal massive gauge bosons.
This parameter space region of constant cross section is where
additional kinematic features are necessary to get more precise
information on the individual couplings $a_{L,T}$.\bigskip

\begin{figure}[b!]
\includegraphics[width= 0.43 \textwidth]{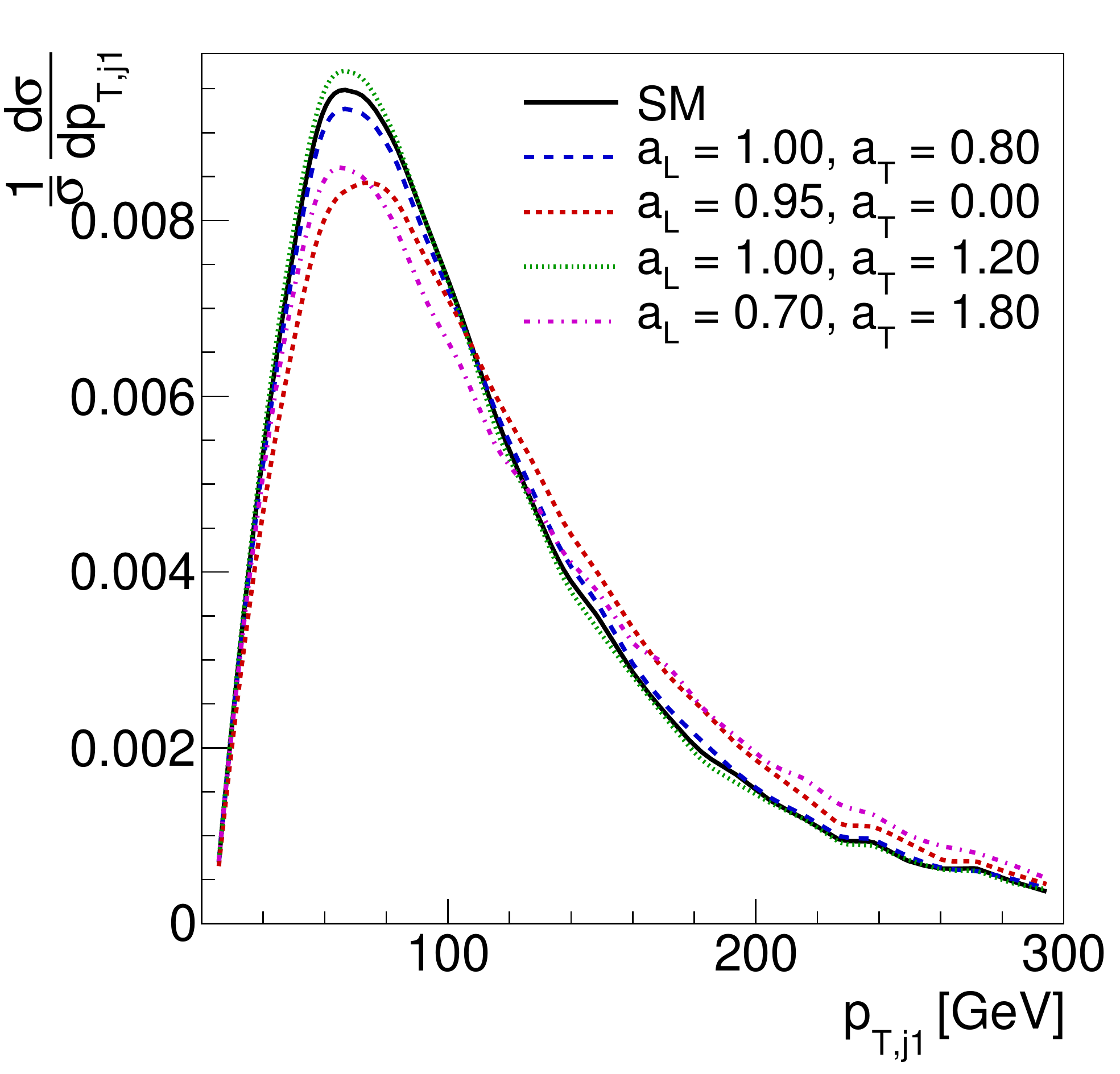}
\hspace*{0.05\textwidth}
\includegraphics[width= 0.47 \textwidth]{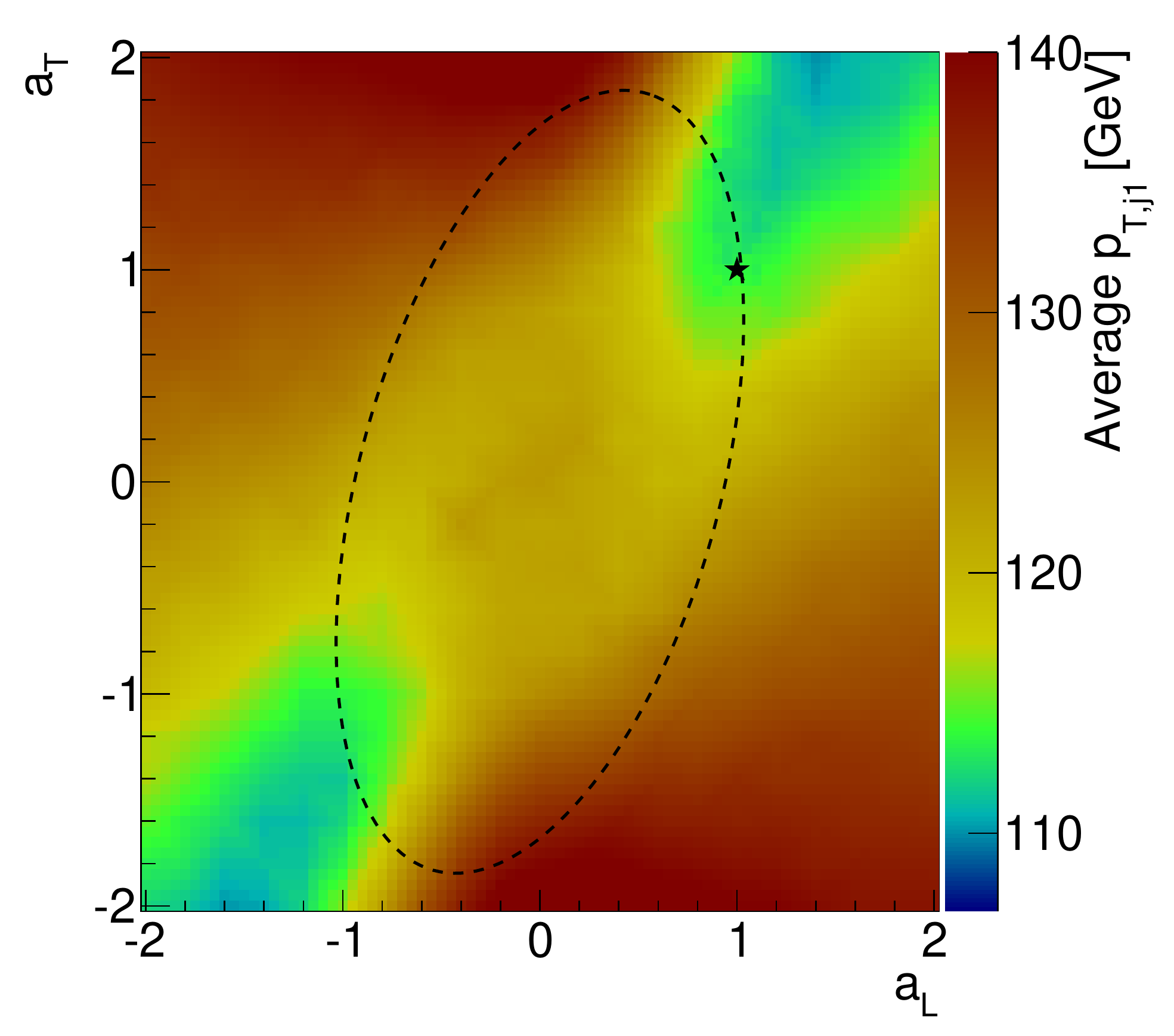}\
\caption{Left: normalized leading $p_{T,j}$ distribution for the a few
  example points, all giving a Standard Model rate.  Right: average leading
  $p_{T,j}$ as a function of the couplings. The ellipse indicates a
  constant rate following Figure~\ref{fig:XSec}.}
\label{fig:JetPt}
\end{figure}

\paragraph{Transverse jet momenta:}
As argued in Section~\ref{sec:wbf} we next focus on the kinematic
features of the two tagging jets. Motivated by the effective $W$
approximation we expect the transverse momentum of both jets to be
sensitive to the Higgs--gauge coupling structure. In the left panel of
Figure~\ref{fig:JetPt} we show distributions of the transverse
momentum of the leading jet for the Standard Model and four additional
parameter points, all giving the same cross section.  In the right
panel we show the average transverse momentum of the leading jet as a
function of the location in parameter space. An analysis of the subleading jet
shows a similar, but slightly less pronounced behavior.

As shown in Figure~\ref{fig:JetPt}, a deviation from the Standard
Model typically leads to a shift to larger transverse momenta of the
tagging jets. While the Standard Model does not mark the $(a_L,a_T)$
point with the minimal average $p_{T,j}$, this minimum is not far
away. It is shifted slightly towards larger transverse coupling. The
effect of modified couplings on the $p_{T,j}$ distributions is clearly
visible, but not huge.  Moreover, it is not clear how well these
signatures survive hadronization, jet clustering and detector effects.

\begin{figure}
\includegraphics[width= 0.45 \textwidth]{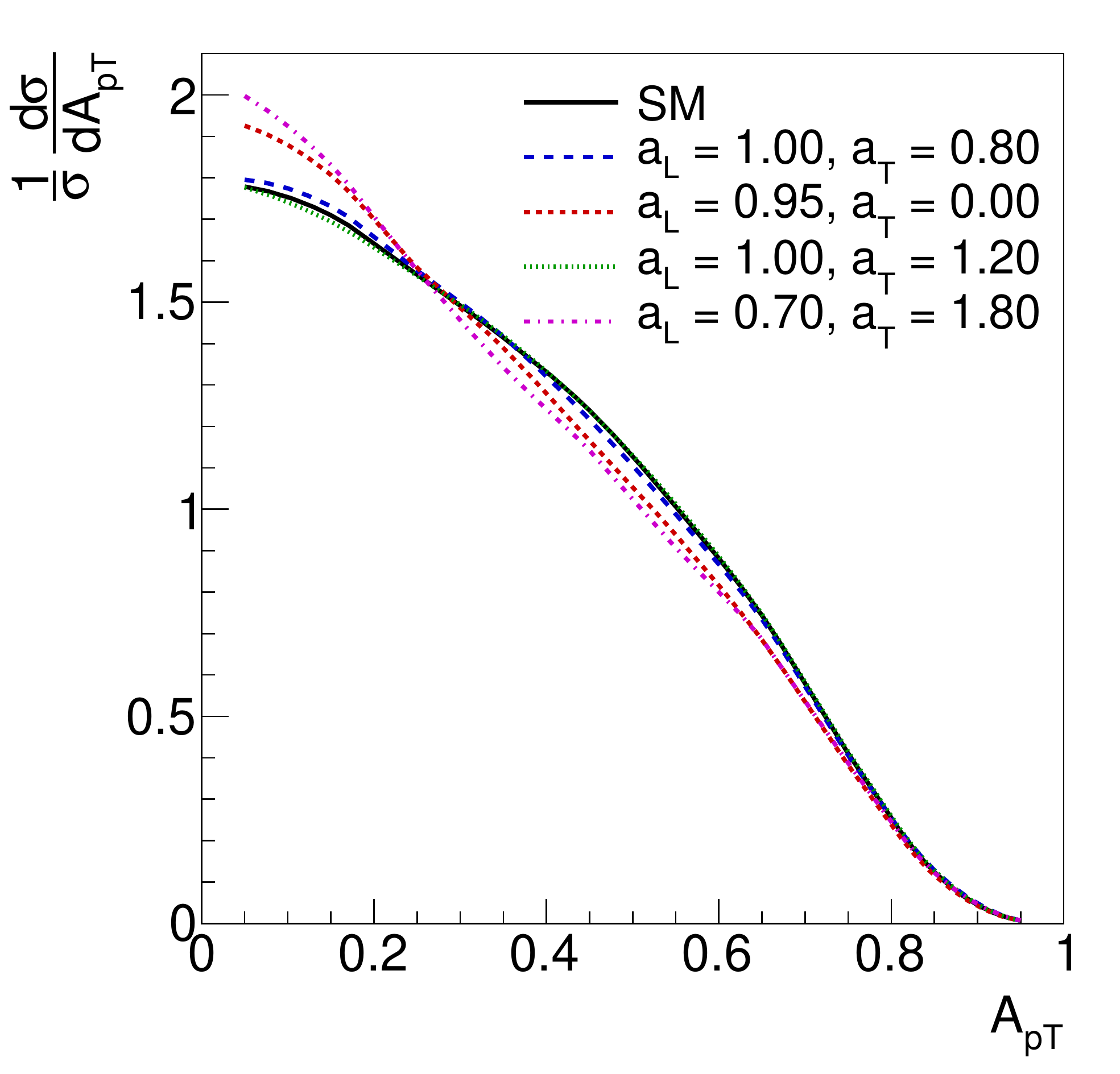}
\hspace*{0.05\textwidth}
\includegraphics[width= 0.45 \textwidth]{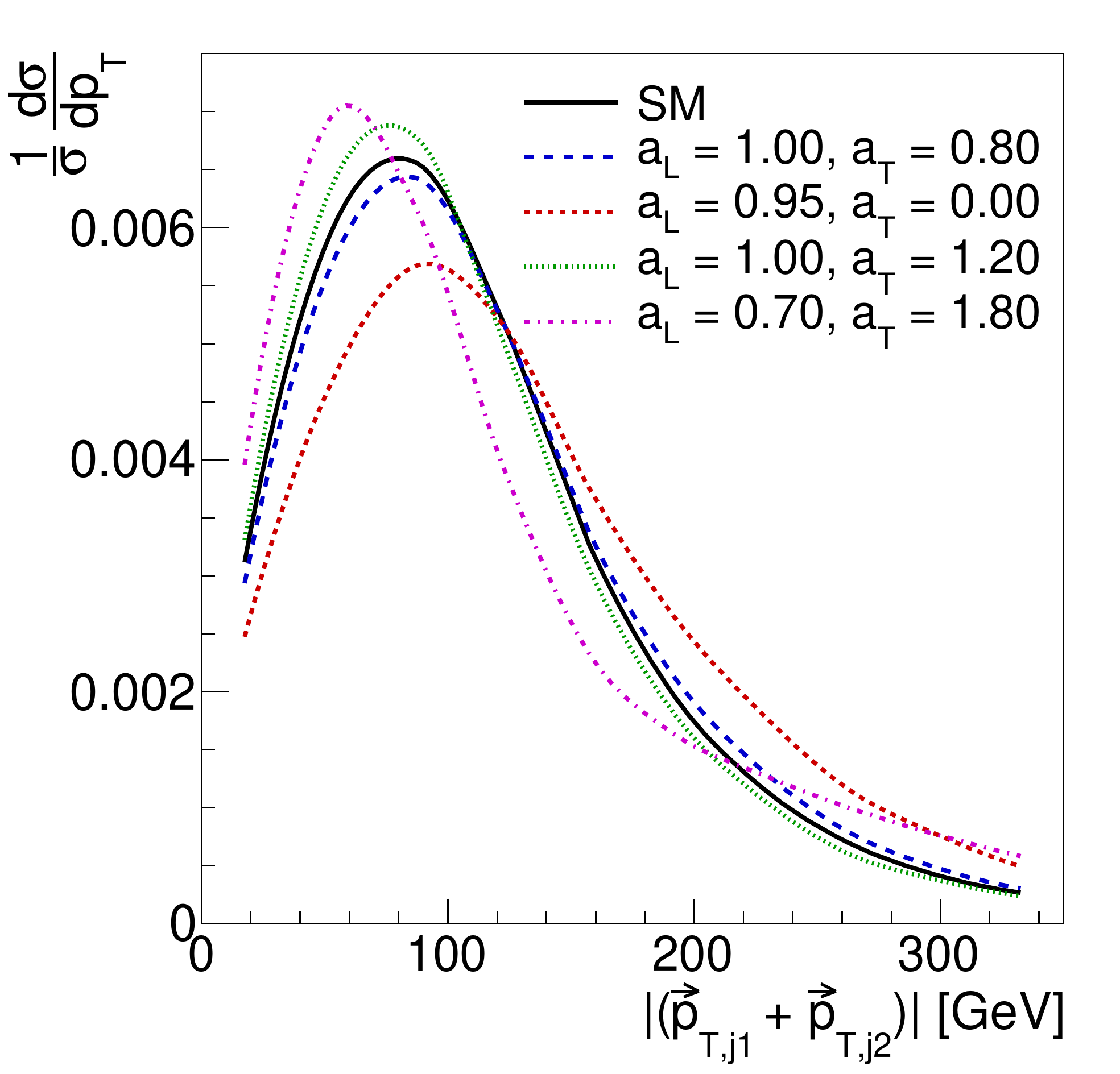}\
\caption{Left: normalized distribution of the asymmetry $A_{pT}$
  between the transverse momenta of the two jets, as defined
  in Eq.~\eqref{eq:PtAvgAsymm}, for a few
  example points, all giving a Standard Model rate.
  Right: normalized distribution of the vectorial sum of the
  transverse momenta of the jets for the same example points.}
\label{fig:JetPtAlt}
\end{figure}

The individual transverse momenta of the leading jet, $p_{T,j1}$, and that
of the subleading jet, $p_{T,j2}$, are not the only potentially
relevant observables. We also consider a parametrization in terms of
the average transverse momentum and the asymmetry in the
transverse momenta between the two jets,
\begin{alignat}{2}
\bar{p}_T = \frac{p_{T,j1} + p_{T,j2}}{2}
 \qqqquad 
A_{pT} = \frac {(p_{T,j1} - p_{T,j2})} {(p_{T,j1} + p_{T,j2})} \; .
\label{eq:PtAvgAsymm}
\end{alignat}
In particular, we expect the asymmetry to be robust under
systematic uncertainties that affect the $p_T$ measurements of both jets alike.

We find that $\bar{p}_T$ behaves very similar to the transverse
momentum of the leading jet: deviations from the Standard Model tend to shift
its distribution to larger values.
In the left panel of Figure~\ref{fig:JetPtAlt} we give distributions
of $A_{pT}$ for the same example points as before. A modification
of the Higgs--gauge couplings slightly favors lower values of $A_{pT}$.
So by moving away from the Standard Model in parameter space, both
jets gain transverse momentum, and the subleading jet increases
by a larger factor than the leading jet.

We also investigate the vectorial sum and difference of the two
transverse jet momenta. Unlike the quantities discussed above,
these observables depend on the angular correlation between the
two jets. The distribution of the vectorial sum is shown in the
right panel of Figure~\ref{fig:JetPtAlt}. For $a_L > a_T$, this
quantity is shifted to larger values, while a more transverse
Higgs--gauge coupling structure favors lower scales.
The vectorial difference changes less strongly with the parameters
and shows the opposite behavior: a shift to more longitudinal couplings
reduces the observable slightly, while a more transverse coupling
structure increases this difference a little.\bigskip

\begin{figure}[t]
\includegraphics[width= 0.43 \textwidth]{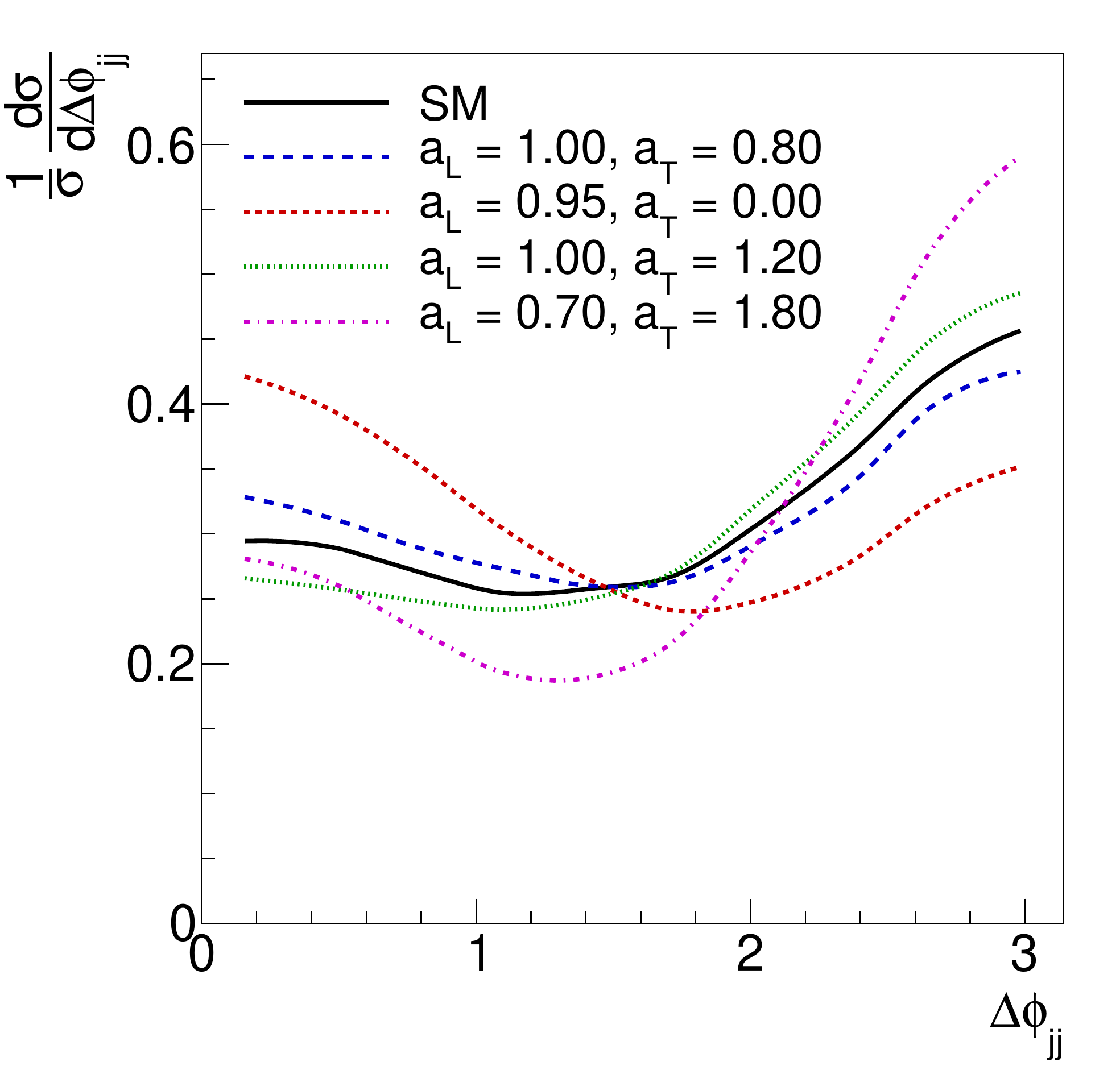}
\hspace*{0.05\textwidth}
\includegraphics[width= 0.47 \textwidth]{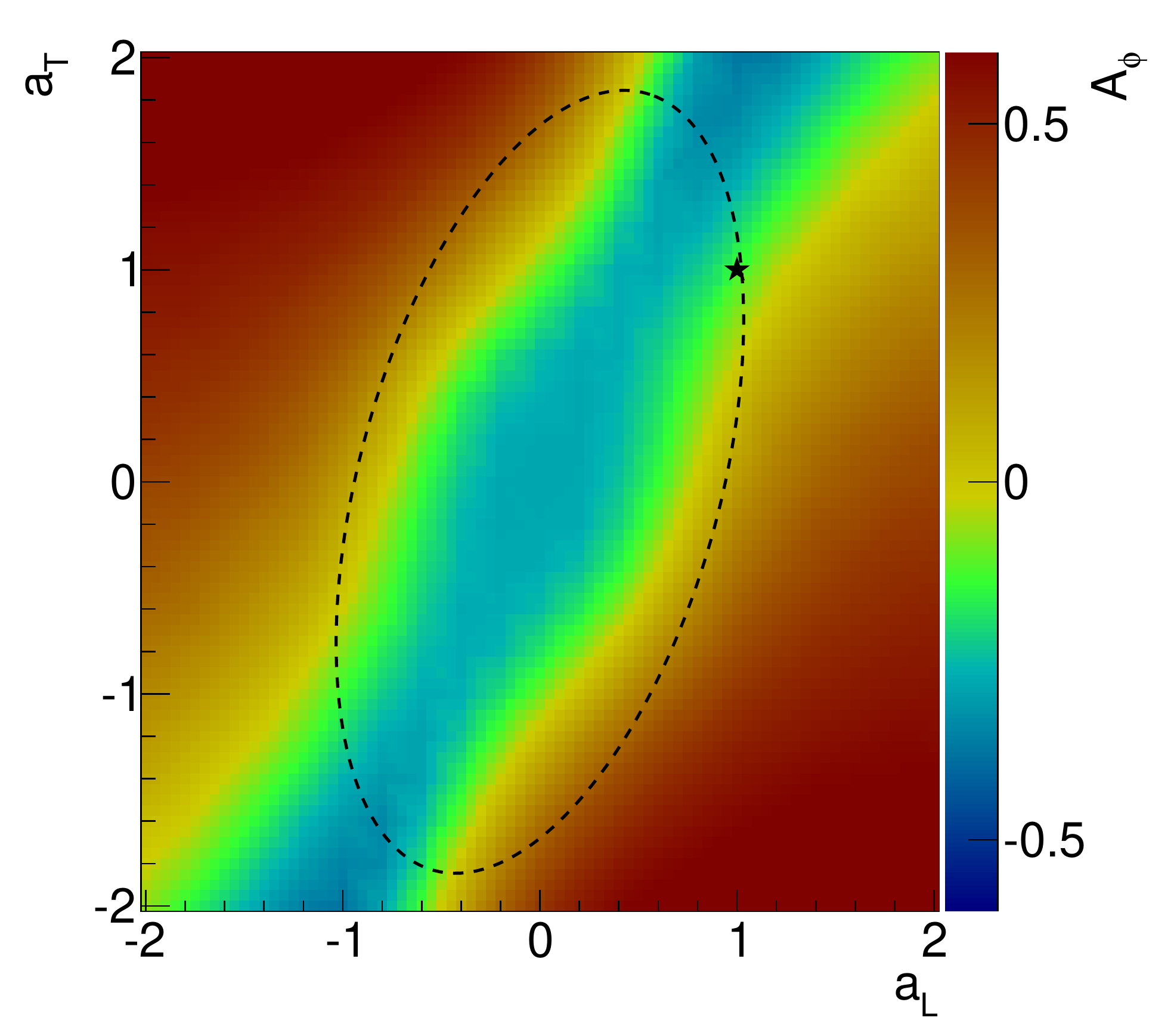}
\caption{Left: normalized $\Delta \phi_{jj}$ distribution for the a
  few example points, all giving a Standard Model rate.
 Right: the asymmetry $A_\phi$ as defined in Eq.~\eqref{eq:asymmetry}
  as a function of the couplings. 
  The
  ellipse indicates a constant rate following Figure~\ref{fig:XSec}.}
\label{fig:PhiJJ}
\end{figure}

\paragraph{Azimuthal angle:}
In addition to the transverse momenta we can also study angular
correlations between the two tagging jets. In the left panel of
Figure~\ref{fig:PhiJJ} we show the distribution of $\Delta
\phi_{jj}$~\cite{delta_phi} for the same parameter points as
before.
It turns out that Higgs--gauge couplings different from the Standard
Model leave a clear signature. Unlike for the transverse momenta, the
Standard Model does not lead to a particular distribution: with $a_L >
a_T$, the jets tend to be more collinear in the transverse
plane. Conversely, a small relative increase of the transverse
coupling favors back-to-back geometries.

In order to quantify this effect in a way that minimizes systematic uncertainties,
we define the asymmetry~\cite{delta_phi_orig}
\begin{equation}
  A_\phi = \frac {\sigma(\Delta \phi_{jj} < \frac \pi 2) - \sigma(\Delta \phi_{jj} > \frac \pi 2)}
  {\sigma(\Delta \phi_{jj} < \frac \pi 2) + \sigma(\Delta \phi_{jj} > \frac \pi 2)} \; .
  \label{eq:asymmetry}
\end{equation}
We give the distribution of $A_\phi$ over the $(a_L, a_T)$ parameter space
in the right panel of Figure~\ref{fig:PhiJJ}. It shows the same behavior as
expected from an analysis of the full $\Delta \phi_{jj}$ distributions.
Deviations from the SM Higgs--gauge coupling are easily visible in $A_\phi$.

The discrimination power of the $p_T$ distributions and of the angular
correlation between the tagging jets is of similar constraining power,
but in an orthogonal direction to the information encoded in the cross
section. A combination of rate measurements with kinematic observables
will therefore efficiently restrict the parameter space consistent
with data.\bigskip

We can also test whether the tagging jet properties are correlated
with the polarization of the final gauge bosons~\cite{tao}. We find
that for the pure Higgs signal they are entirely uncorrelated,
reflecting the scalar nature of the Higgs boson where the decay mode
knows nothing about its production. Including the backgrounds, a
correlation emerges, caused by the fact that final-state longitudinal
$WW$ pairs are more likely to stem from a signal events, while
transverse bosons have a larger probability to come from a background
interaction.\bigskip

\paragraph{Other observables:}
So far, we have restricted our analysis to observables describing the
jet kinematics in the transverse plane, and found them to be sensitive to the
Higgs--gauge coupling structure. We have also analyzed the distribution
of quantities that are (mostly) sensitive to the longitudinal jet momentum.
These include the jet energies, the invariant mass between the two
jets, as well as their separation in pseudorapidity. We find that all of these
observables are much less sensitive to the structure of the Higgs--gauge
sector than the purely transverse quantities discussed above.
This does not come as a surprise: the longitudinal momentum
of the final-state quarks in the WBF topology is dictated by the incoming
quarks, which at the LHC set a much larger scale than that describing the hard
$WW\to H$ process.
Since including these variables does not improve the
significance of our findings, we limit our analysis to the transverse
jet momenta and the angular correlation between the jets in the
transverse plane.\bigskip

\begin{table}
    \begin{ruledtabular}
      \begin{tabular}{l@{\qquad}c@{\quad}c}
      Observables & Limit on $a_L$ & Limit on $a_T$\\
      \hline
      $\sigma$  & ($\leq$ 1.07) & ($\leq$ 1.97) \\
      $\sigma$, $p_{T,j1}$ & (0.76\;--\;1.08) & (0.25\;--\;1.79) \\
      $\sigma$, $p_{T,j1}$, $p_{T,j2}$ & (0.82\;--\;1.08) & (0.56\;--\;1.68) \\
      $\sigma$, $\bar{p}_T$ & (0.79\;--\;1.07) & (0.41\;--\;1.73) \\
      $\sigma$, $A_{pT}$ & (0.65\;--\;1.08) & ($\leq$ 1.86) \\
      $\sigma$, $\bar{p}_T$, $A_{pT}$ & (0.78\;--\;1.09) & (0.39\;--\;1.73) \\
      $\sigma$, $\Delta\phi_{jj}$ & (0.49\;--\;0.54) and  (0.95\;--\;1.06) & (0.83\;--\;1.17) and (1.89\;--\;1.94) \\
      $\sigma$, $A_{\phi}$ & (0.52\;--\;0.64) and  (0.94\;--\;1.06) & (0.82\;--\;1.15) and (1.77\;--\;2.00) \\
      $\sigma$, $|(\vec{p}_{T,j1} + \vec{p}_{T,j1})|$ & (0.93\;--\;1.06) & (0.66\;--\;1.28) \\
      $\sigma$, $|(\vec{p}_{T,j1} - \vec{p}_{T,j1})|$ & ($\leq$ 0.61) and  (0.85\;--\;1.08) & ($\leq$ 1.96) \\
      $\sigma$, $p_{T,j1}$, $p_{T,j2}$, $\Delta\phi_{jj}$ & (0.92\;--\;1.08) & (0.82\;--\;1.19) \\
      $\sigma$, $p_{T,j1}$, $p_{T,j2}$, $A_{\phi}$ & (0.92\;--\;1.08) & (0.80\;--\;1.18) \\
    \end{tabular}
  \end{ruledtabular}
  \caption{Limits on $a_{L,T} \in [0,2]$ in the absence of a signal
    based on different observables. The limits are given at 95$\%$ CL
    assuming $300~\ifb$. For the limit on $a_L$, $a_T$ is allowed
    to float freely, and vice versa.}
  \label{tbl:Limits_observables}
\end{table}

\paragraph{Combination:}
In a next step, we estimate the sensitivity in the $(a_L,a_T)$ plane
when measuring these kinematical features. This will give us the
parameter space which can be probed during the upcoming LHC run.
We generate a number of toy data samples based on the Standard Model,
representing an integrated luminosity of $300~\ifb$. Each of these
toy data samples is compared to the samples based on our simple model.
We determine the significance of deviations in the Higgs cross section
and the asymmetry $A_\phi$ by calculating the probability density
functions of these quantities for each parameter point of our simple
model.  For the other kinematic observables discussed above, we
measure the significance of deviations by performing $\chi^2$ tests on
the normalized distributions.
From these tests we extract the median $p$-value. If it is below $0.05$,
the parameter point $(a_L, a_T)$ in question is expected to be
excluded at $95\%$~CL in the absence of a signal. The results from
certain sets of observables are statistically independent and can be
combined correspondingly.

In Table~\ref{tbl:Limits_observables} we give the limits on the parameters
$a_L$ and $a_T$ obtained using different combinations of observables.
In general, observables including the angular correlation between the jets
give stronger constraints than those based only on transverse momenta.
However, an analysis based only on the $\Delta\phi_{jj}$ distribution or
only on the asymmetry $A_\phi$ leads to a binary ambiguity: there is a blind spot
in the parameter space around the parameters $a_L \approx 0.6$, $a_T \approx 1.8$.
This region shows the same rate and angular correlation between the jets
as the Standard Model. Including the transverse jet momentum in the analysis removes this
ambiguity.
All in all, a combination of the cross section with the transverse momenta of the leading
and subleading jet as well as the asymmetry $A_\phi$ yields the strongest
constraints on the parameters of our simple model.\bigskip

\begin{figure}
\includegraphics[width= 0.43 \textwidth]{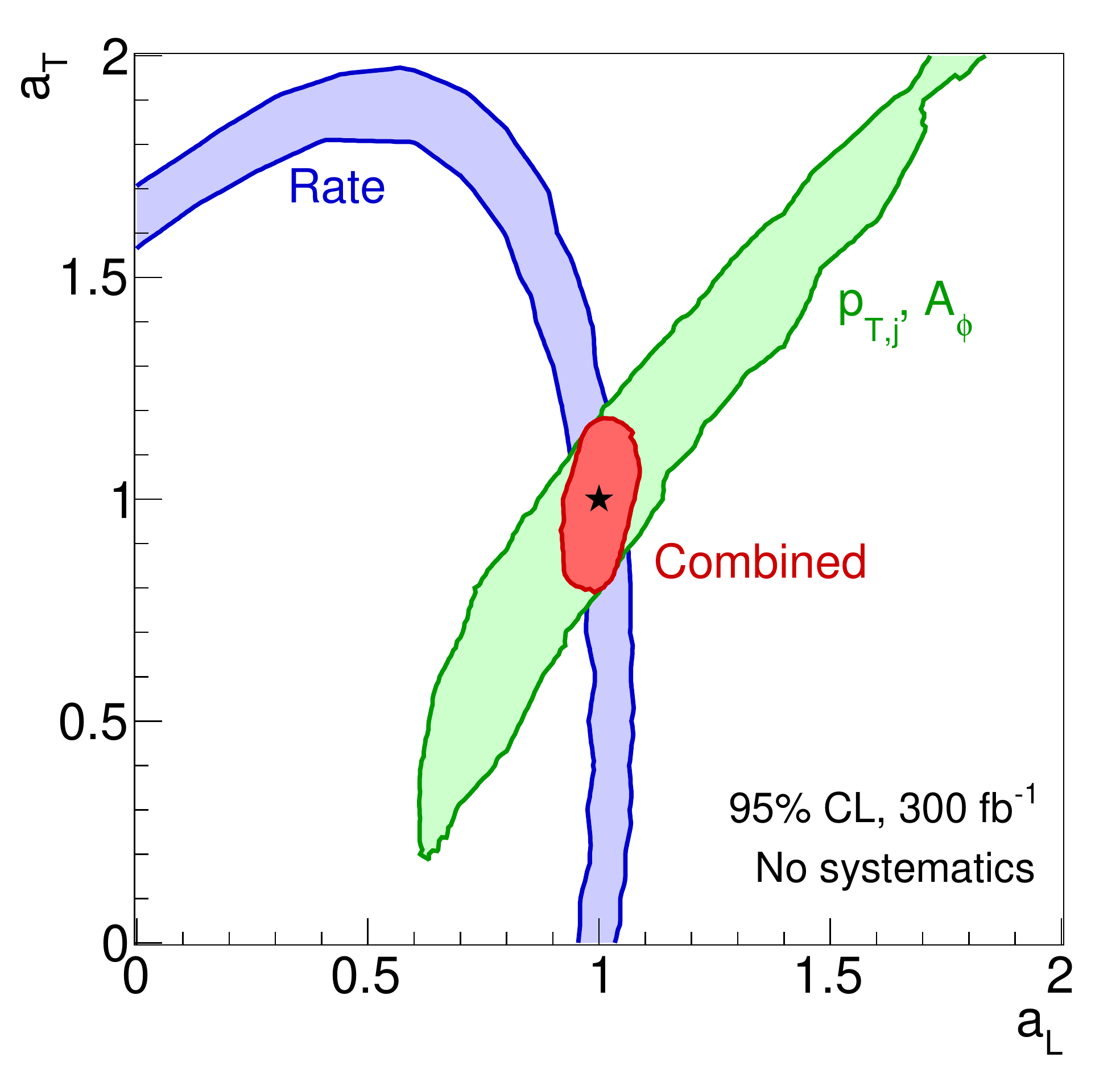}
\hspace*{0.05\textwidth}
\includegraphics[width= 0.43 \textwidth]{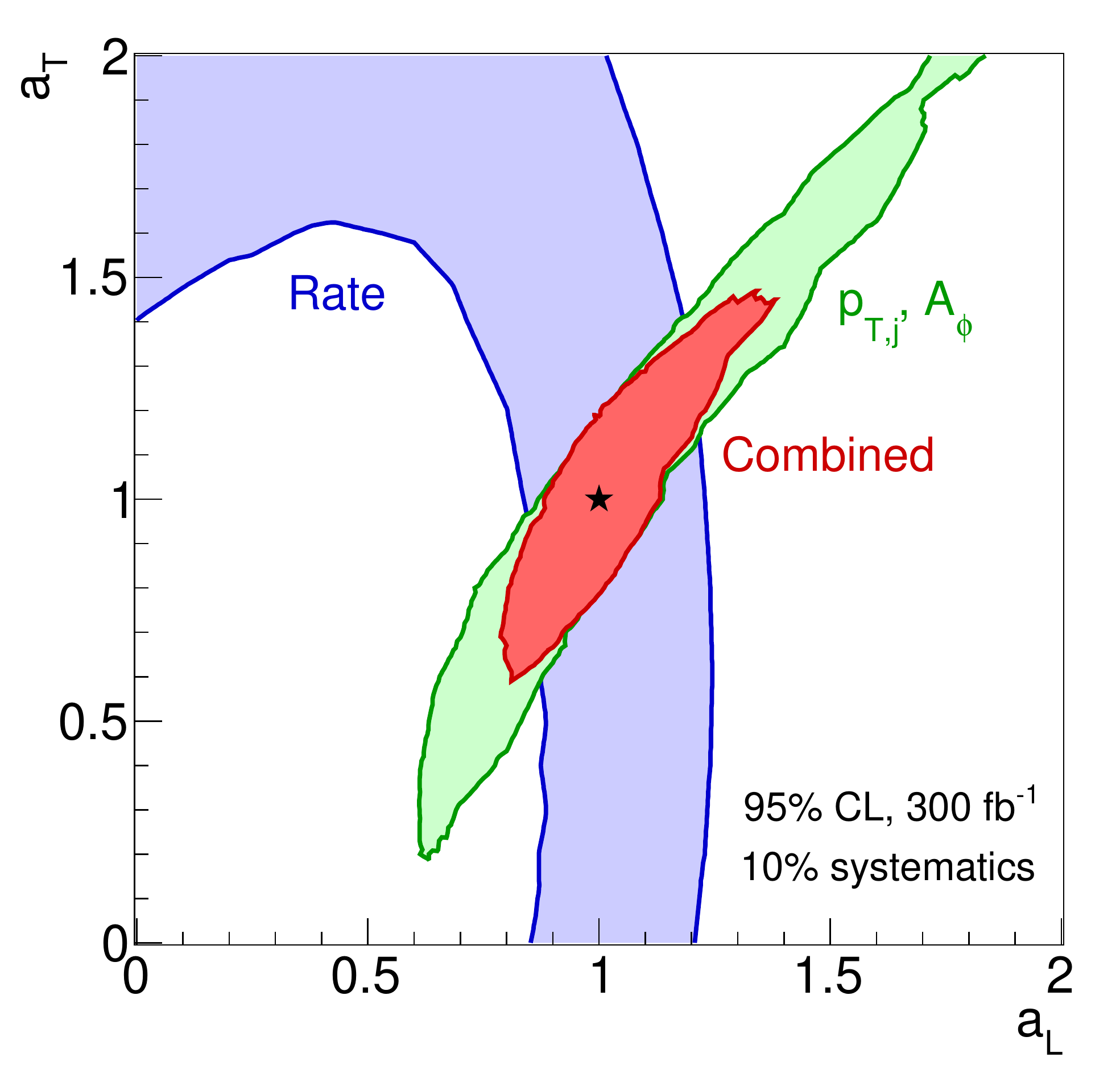}
\caption{Expected exclusion regions at 95$\%$ CL after $300~\ifb$
  of data in the absence of a signal. Note that these are results
  calculated at parton level and not including any systematic
  uncertainties. In the right panel we show the same result including
  an additional 10\% uncertainty on the Higgs production and decay
  rate.}
\label{fig:Exclusion}
\end{figure}

In Figure~\ref{fig:Exclusion} we show the expected exclusion regions
for these observables. With an integrated luminosity of $300~\ifb$
they can be used to exclude most of the $(a_L,a_T)$ plane.
At least on parton level and disregarding
systematic uncertainties, the coupling factor $a_L$ should be
measurable at $\ord(10 \%)$, while the transverse coupling
should be measurable at $\ord(20\%)$. The mirrored solution
with $a_L \approx -1$, $a_T \approx -1$ cannot be excluded in this
analysis. However, the $H\to\gamma \gamma$ decay, mediated by a top
loop and a $W$ loop, is easily sensitive to this sign change.

While we neglect theoretical and systematic uncertainties, we also
show the results of the same analysis after including a hypothetical
uncertainty of $\pm 10\%$ on the Higgs production and decay rate.
At this level of analysis it is simply not clear how large the different
uncertainties are. This will largely depend on the central jet veto or
alternative analysis~\cite{cjv} and the associated theoretical
uncertainties.

\section{Choice of reference frame}
\label{sec:frame}

\begin{figure}[t]
  \includegraphics[width=0.45 \textwidth]{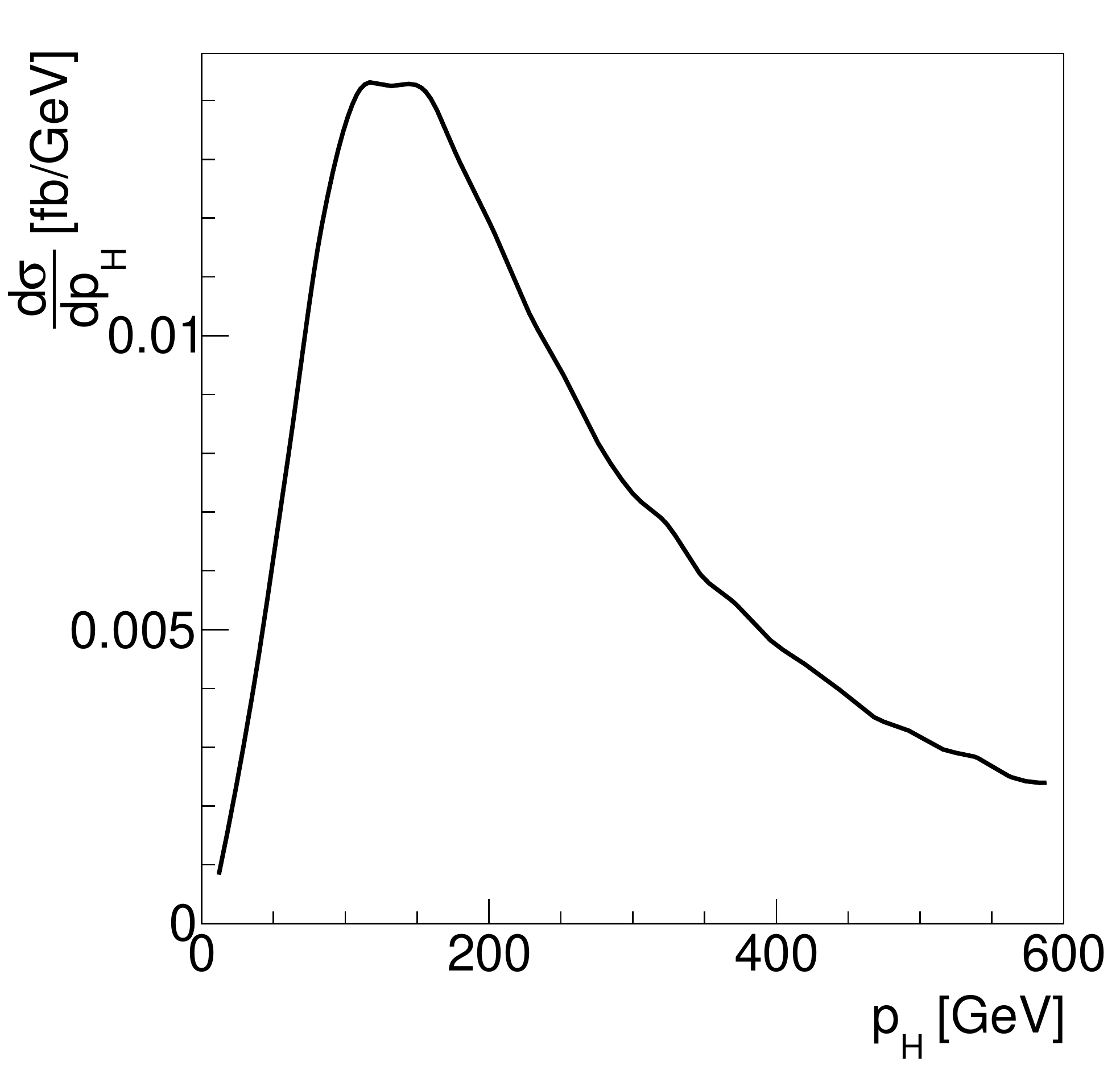}
  \hspace*{0.05\textwidth}
  \includegraphics[width=0.45 \textwidth]{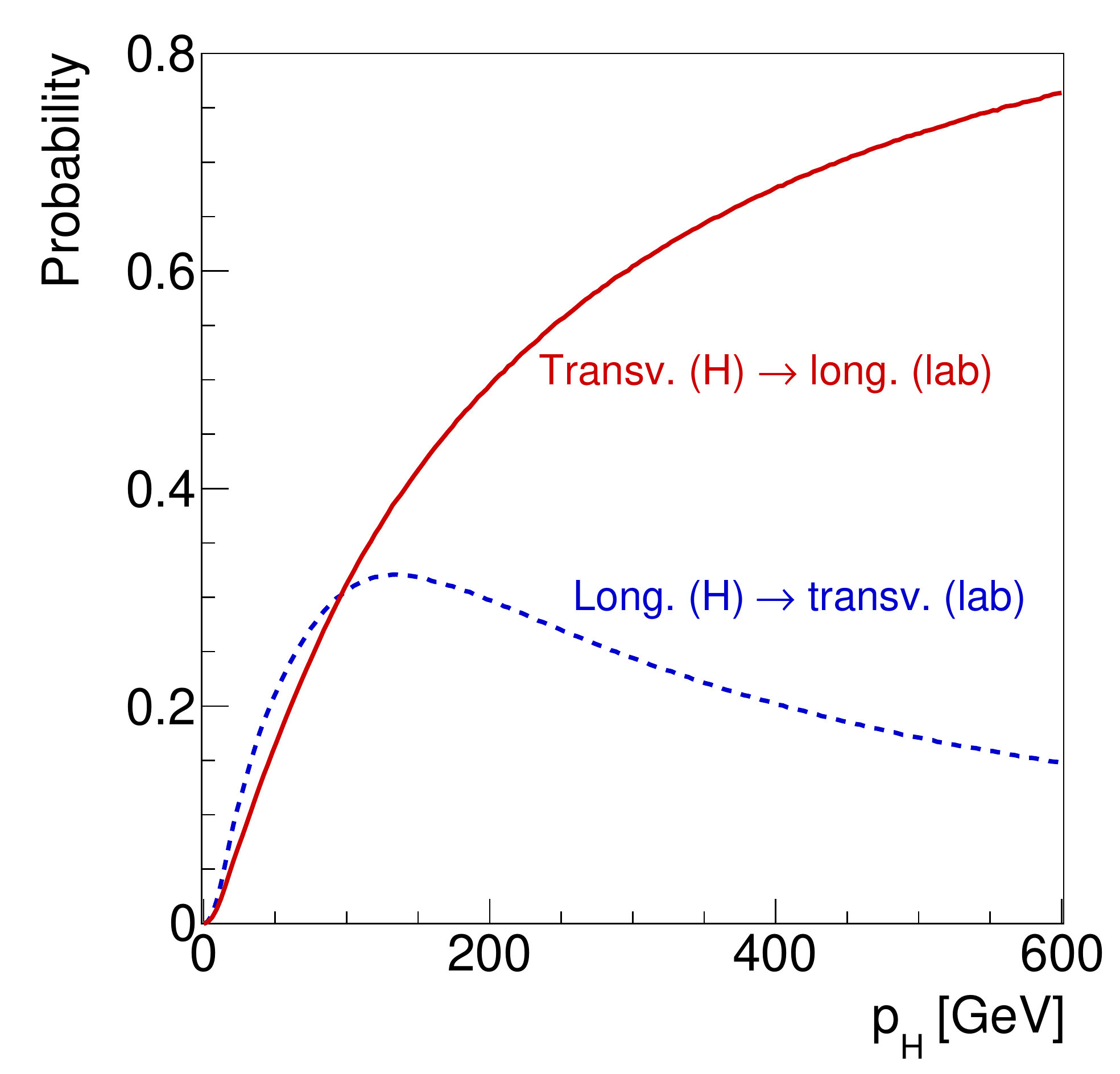}
  \caption{Left: distribution of the Higgs momentum in the SM after
    our selection cuts. Right: probabilities for a change in
    polarization when going from the Higgs rest frame to the
    laboratory frame as a function of the Higgs momentum.}
  \label{fig:frame}
\end{figure}

In the setup described above we define the polarization of the virtual
gauge bosons in the Higgs rest frame. This leads us to results fully
consistent with the expectations from the effective $W$
approximation. However, while the definition of polarizations requires
a reference frame, in the absence of a $V$ rest frame there remains a
choice of frames. In this section we will briefly
review the effect of a different choice.

To link the different reference frames we have to understand how
polarizations change during boosts.  The tagging jet distributions
probe the $HVV$ vertex in the weak--boson--fusion production of an
on-shell Higgs boson.  Here, we will for simplicity consider the Higgs
decay to an on-shell and an off-shell $V$ boson instead.  The
polarization of the $V$ bosons in the Higgs rest frame changes when we
apply a boost into another frame with a finite Higgs momentum
$\vec{p}_H$ in that new frame. The probability of a transverse boson
becoming longitudinal or vice versa depends on the size of the boost
as well as the angle between the boost direction and the $V$ momenta,
where in our brief discussion we average over the relative angle.  In
the left panel of Figure~\ref{fig:frame} we give the distribution of
the Higgs momentum for our reference process in the laboratory frame.
It ranges around $100$\;--\;$300$~GeV. In the rest frame of the two
colliding partons the typical Higgs momentum is slightly
reduced.

Given this momentum range we calculate the probability for a
transverse $V$ in the Higgs rest frame to become longitudinal after
the boost, and vice versa. We give the results in the right panel of
Figure~\ref{fig:frame}.  For Higgs momenta around 200~GeV, corresponding
to a $\gamma \sim 2$, the probability for a change in polarization
reaches $\ord(50\%)$; the polarizations are essentially
randomized. Any clear effect in the Higgs rest frame is washed out in
any frame with such a relative boost.\bigskip

In the following we will show that the $VV$ rest frame or
equivalently the Higgs rest frame is well suited to analyze
polarization effects in $VV$ scattering. Moreover, we will see that a
definition in this frame also nicely compares with an approach based
on higher-dimensional operators. Defining the polarizations in the
laboratory frame or the rest frame of the two colliding partons
therefore leads to qualitatively similar, but much less pronounced
findings.

\section{Effective field theory}
\label{sec:dim6}

Complementing our simple model we can parametrize structural
deviations from the Standard Model in terms of higher-dimensional
operators. This has the advantage of being manifestly Lorentz
invariant, so it can serve as a check of our assumption that the
definition of the transverse vs longitudinal reference frame at finite
energies has little impact on our physics results.  

Without adding new operator structures to the Standard Model
Lagrangian $\lag_\text{SM}$ we can allow for changes in the Higgs
coupling strength to two vector bosons. For more drastic changes, in
particular affecting longitudinal and transverse gauge bosons
differently and hence leading to different tagging jet properties,
we invoke new operator structures.  The question is how the obvious
physical description of gauge boson polarizations used in the previous
sections can be phrased in terms of such higher dimensional operators,
\begin{equation}
  \lag_\text{D6} = \lag_\text{SM} + \sum_i \frac{c_i}{\Lambda^2} \ope_i \; .
\end{equation}
As an example we will study two electroweak operators including
the usual Higgs doublet $\phi$,
\begin{alignat}{5}
\ope_{\phi,2} = 
\frac{1}{2} \partial_\mu \left( \phi^\dagger \phi \right) \partial^\mu \left( \phi^\dagger \phi \right)
\qquad \text{and} \qquad   
\ope_W &= \left(D_\mu \phi \right)^\dagger {\hat W}^{\mu \nu} \left( D_\nu \phi \right ) \; ,
\label{eq:OW}
\end{alignat}
following the conventions discussed in
Refs.~\cite{hisz,d6_review,other_bases}. For instance, these operators
arise in strongly interacting Higgs sectors~\cite{silh,composite},
including little Higgs~\cite{little_higgs} and holographic
Higgs~\cite{holographic_higgs} models.\bigskip

First, the operator $\ope_{\phi,2}$ generates a new
contribution to the Higgs kinetic term $\partial_\mu H \partial^\mu
H$. A field redefinition $H \to H / \sqrt{1 + c_{\phi 2} v^2
  /\Lambda^2}$ restores (most of) the canonical normalization of the
kinetic term~\cite{silh,lecture}, but also introduces a form factor
for every Higgs vertex. Following the definition in
Eq.~\eqref{eq:lag_pol}, the $HWW$ interaction is modified by the
universal factor 
\begin{equation}
a_L = a_T = 1 - \frac {c_{\phi 2}} 2 \frac{v^2}{\Lambda^2} \; ,
\label{eq:Dim6EquivalentParam1}
\end{equation}
that does not depend on the momenta. It is therefore in one-to-one
correspondence with the $a_L = a_T$ case in our simple model.  Because
$\ope_{\phi,2}$ only shifts Higgs interactions universally it is
hardly constrained by electroweak precision measurements and the
measurement of triple gauge vertices.\bigskip

For the operator $\ope_W$ the situation is more interesting.
It modifies the interaction between the weak gauge bosons and the
Higgs boson, but there is no term of order $\phi^3$ or higher
contributing to the Higgs--Goldstone couplings. At large energies it
therefore shifts the transverse Higgs--gauge coupling, but not its
longitudinal counterpart.  While the Higgs resonance hardly
constitutes a high-energy limit, we expect the effect of
$\ope_W$ on the longitudinal coupling $a_L$ to be suppressed as
compared to $a_T$.

In unitary gauge the operator $\ope_W$ yields two types of
corrections to the $HWW$ vertex,
\vskip0.5cm
\begin{equation}
  \parbox{12mm}{
    \begin{fmfgraph*}(30,30)
      \fmfleft{i}
      \fmfright{o2,o1}
      \fmflabel{$H$}{i}
      \fmflabel{$W^+_\mu$}{o1}
      \fmflabel{$W^-_\nu$}{o2}
      \fmf{dashes}{i,v}
      \fmf{boson}{v,o1}
      \fmf{boson}{v,o2}
    \end{fmfgraph*}
  }
  \quad = i g m_W  
  \Big[   
  \underbrace{g_{\mu \nu} - g_{\mu \nu} \frac{c_W}{2 \Lambda^2} \left( (p^H p^+) + (p^H p^-) \right)}_{a_{L,T}^{(1)}}
  + \underbrace{\frac{c_W}{2 \Lambda^2} \left( p^H_\mu p^+_\nu + p^-_\mu p^H_\nu \right)}_{a_{L,T}^{(2)}}   
  \Big] \; ,
  \label{eq:OW_HWW}
\end{equation}
\vskip0.5cm
\noindent 
where $p^\pm_\mu$ and $p^H_\mu$ are the incoming momenta of the
$W^\pm$ and the $H$, respectively. The first two terms of
Eq.~\eqref{eq:OW_HWW} correspond to the Standard Model vertex and an
unchanged Lorentz structure with a higher-dimensional coupling
strength modification. For an on-shell Higgs
these two terms give us 
\begin{equation}
 a^{(1)}_L = a^{(1)}_T = 1 + \frac{c_{W}  }{2 \Lambda^2} m_H^2 \; .
\label{eq:Dim6EquivalentParam2}
\end{equation}
The last term in Eq.~\eqref{eq:OW_HWW} features contractions of the
type $(p^H \varepsilon^\pm)$, where $\varepsilon^\pm$ are the
polarization vectors of the $W^\pm$ bosons. In the Higgs rest frame,
this terms vanishes for transverse gauge bosons,
but its contribution to the longitudinal coupling does not.
Unlike in our simple model, this factor is a function of the
$W$ momenta,
\begin{equation}
a^{(2)}_T = 0 \qqqquad 
a^{(2)}_L = \frac{c_W}{\Lambda^2} \; F(p^+,p^-) \; .
\label{eq:Dim6EquivalentParam3}
\end{equation}
The operator $\ope_W$ does not only affect Higgs physics, but
clearly also contributes to the $ZWW$ and $WWW$ interactions. Still,
there are no strong constraints from electroweak precision
measurements of the oblique parameters $S$, $T$ and $U$. This is
because the loop contributions from this operator can  be
balanced by other operators, making this operator a so-called `blind
direction'~\cite{BlindDirection}.  The strongest limits on $c_W$ hence
stem from measurements of triple gauge boson vertices at the
LHC~\cite{Dim6Limits}.\bigskip

\begin{figure}[t]
\includegraphics[width= 0.366 \textwidth,clip,trim=0 0 1cm 0 ]{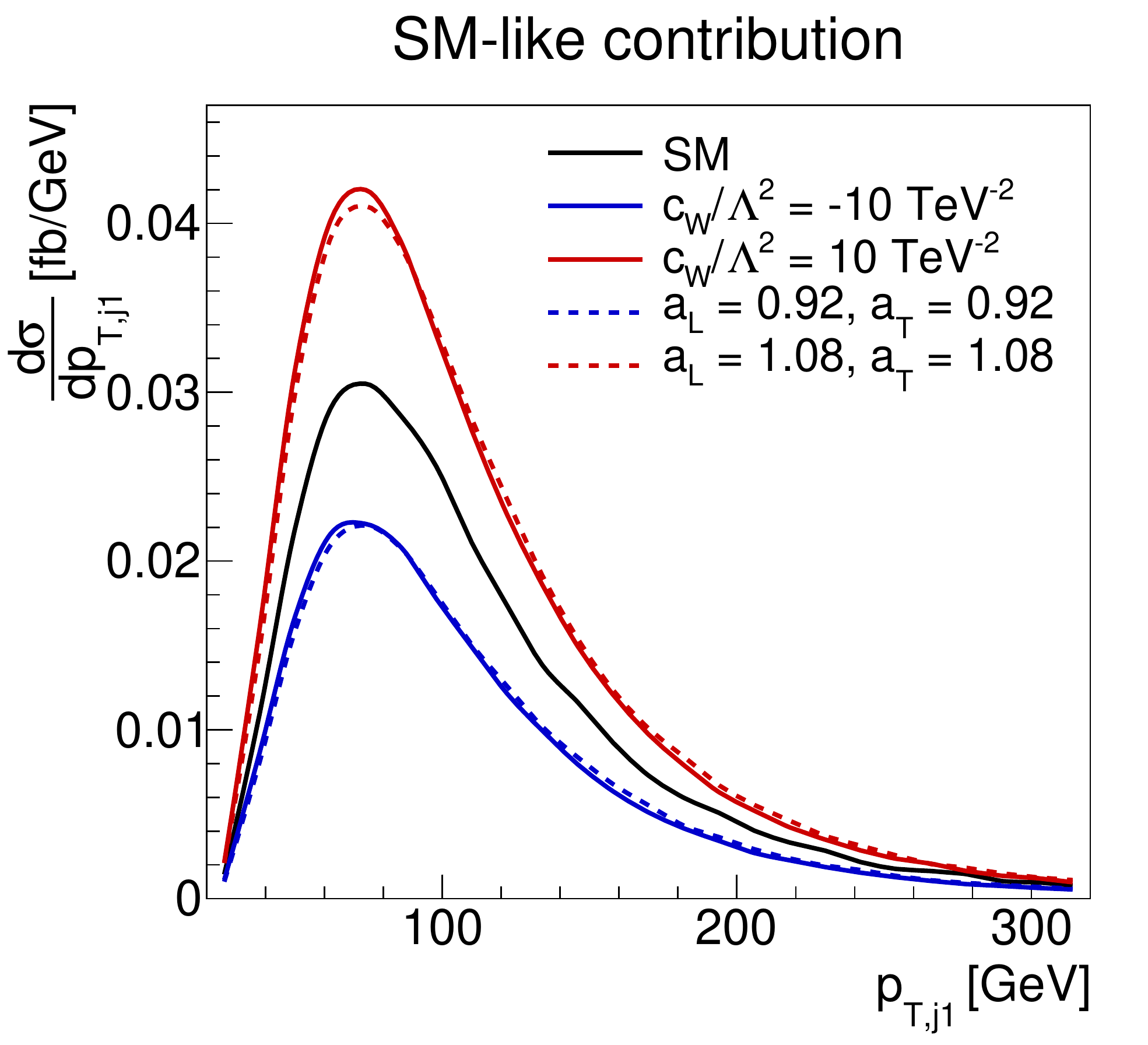}
\includegraphics[width= 0.2965 \textwidth,clip,trim=3.6cm 0 1cm 0 ]{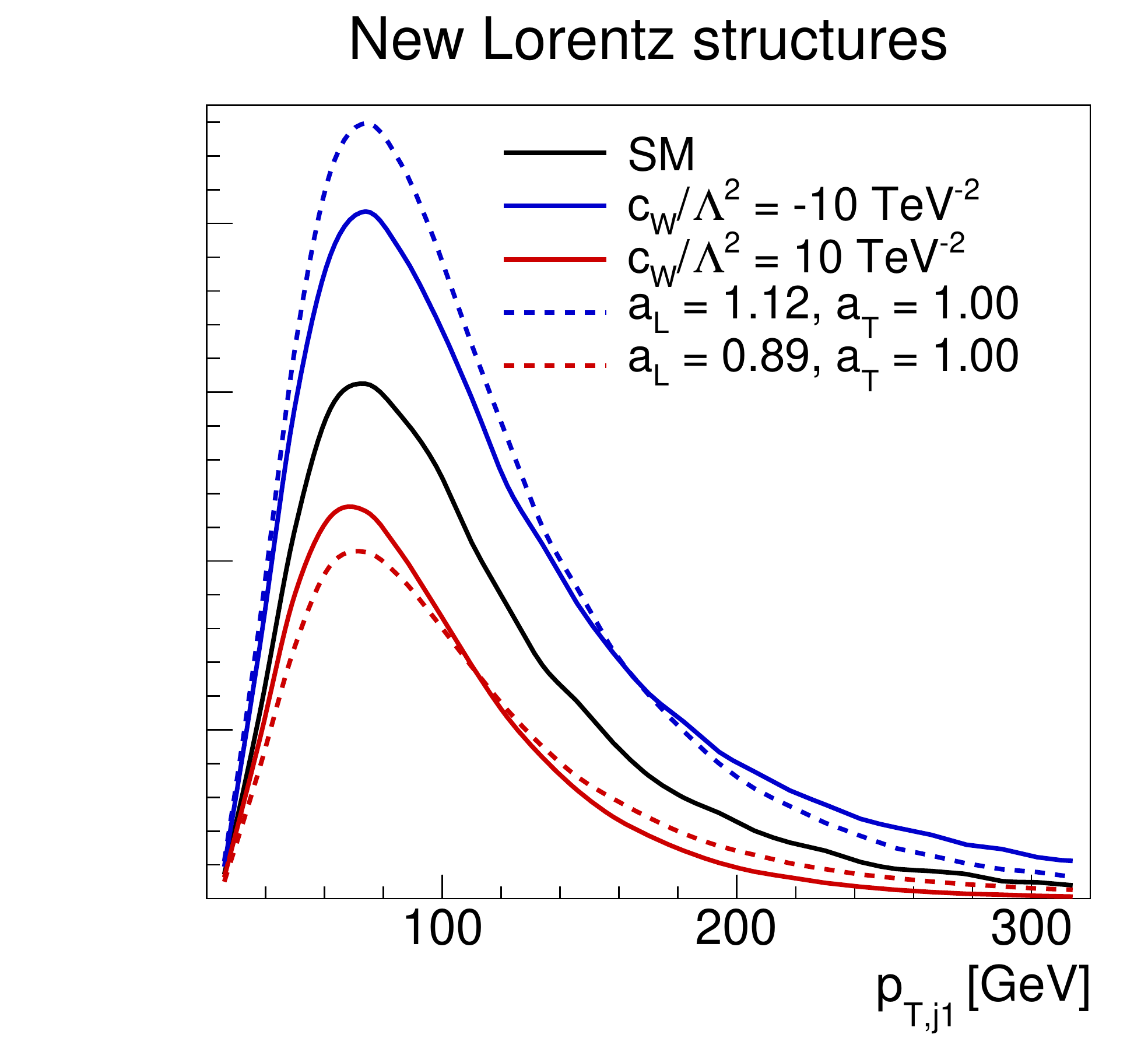}
\includegraphics[width= 0.316 \textwidth,clip,trim=3.6cm 0 0 0 ]{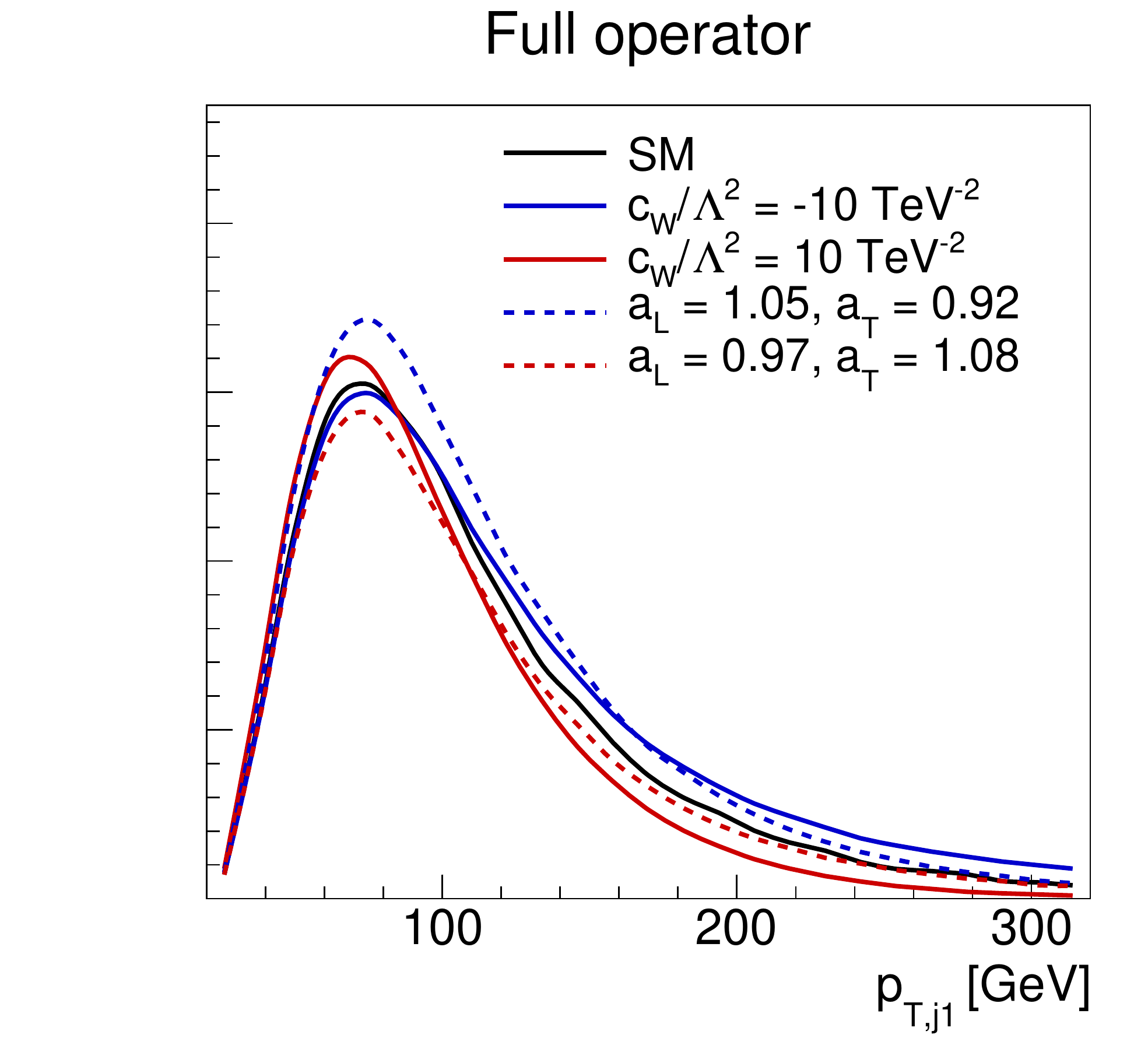} \\
\includegraphics[width= 0.366 \textwidth,clip,trim=0 0 1cm 0.8cm]{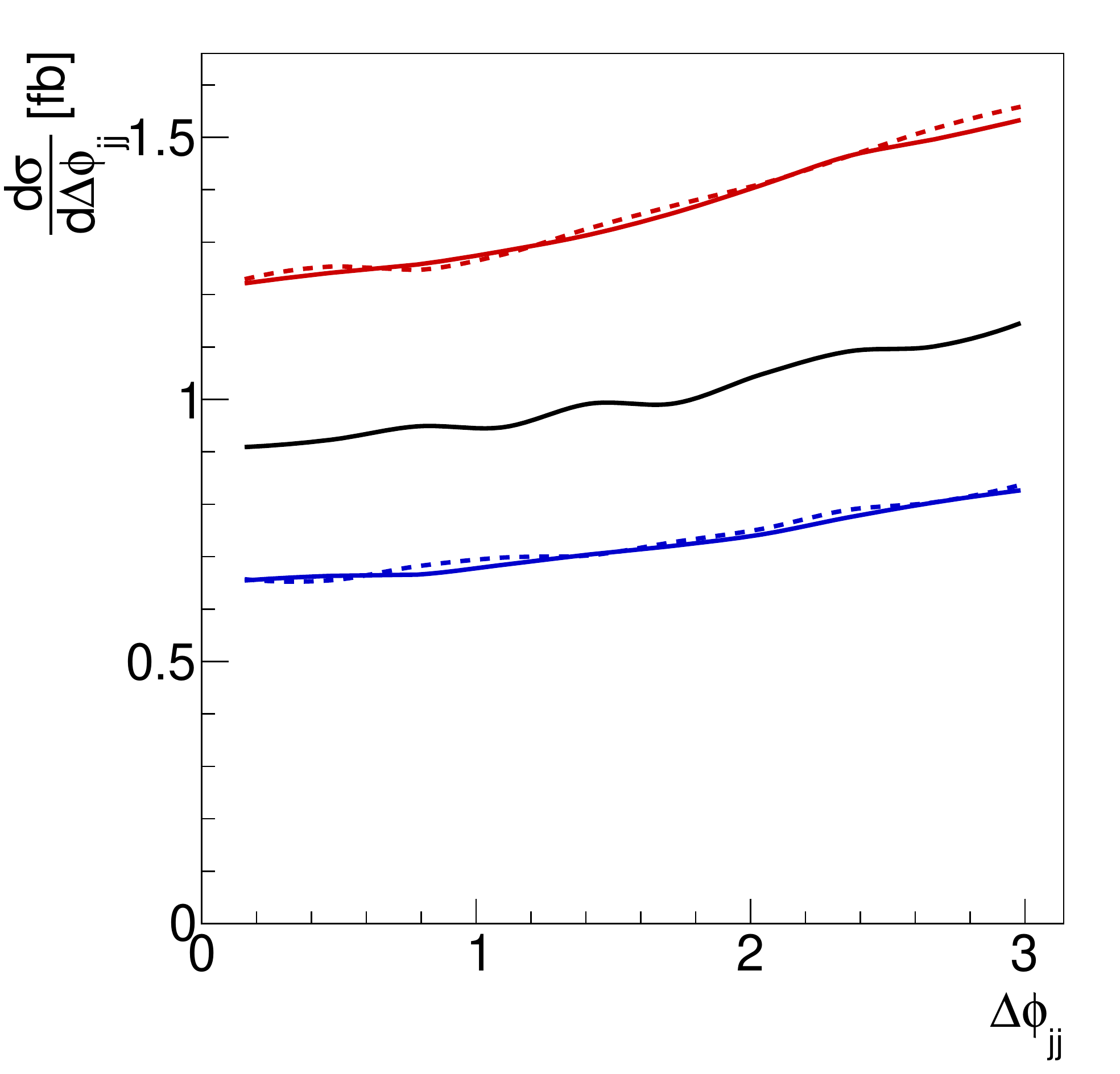}
\includegraphics[width= 0.2965 \textwidth,clip,trim=3.6cm 0 1cm 0.8cm ]{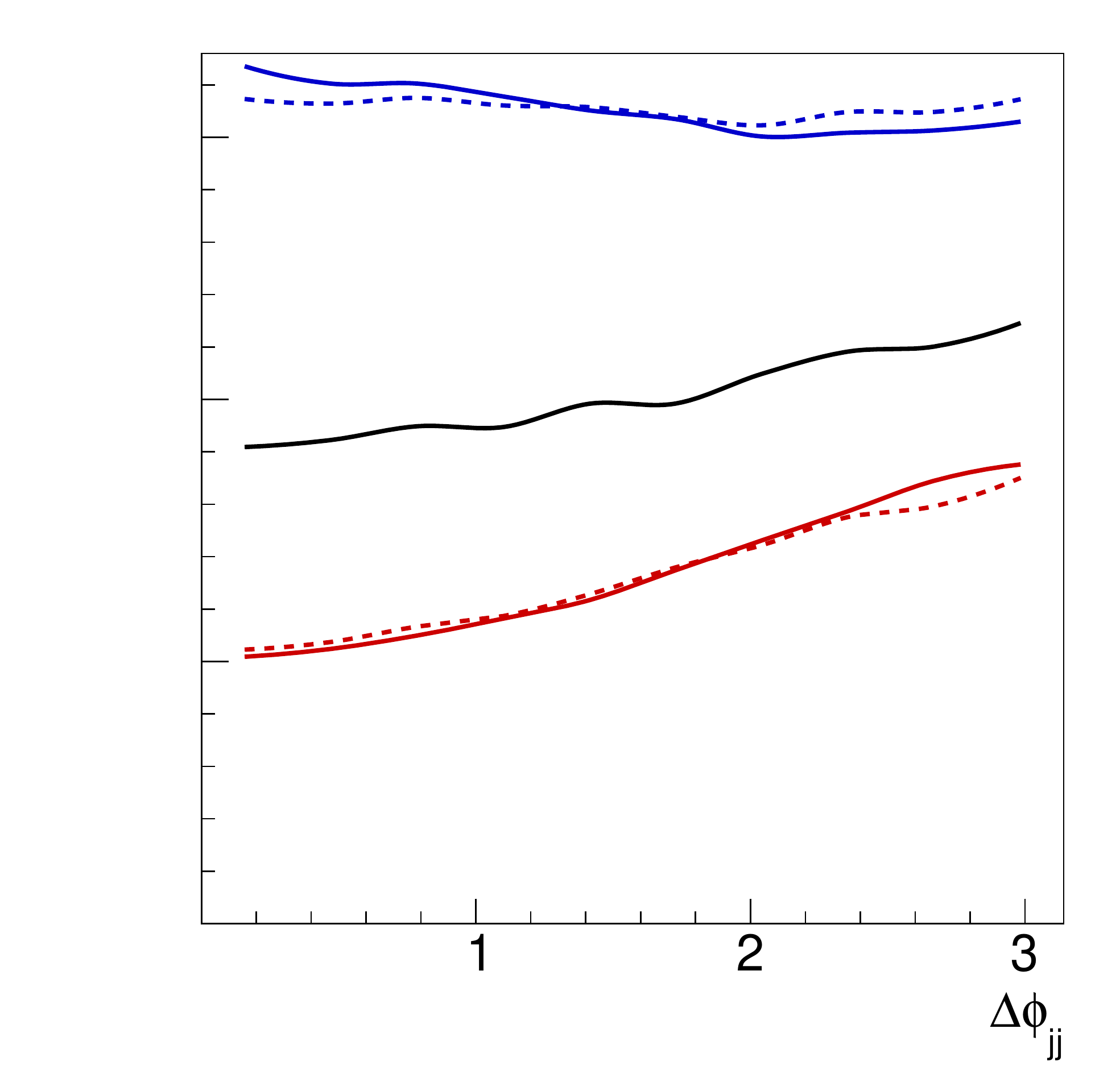}
\includegraphics[width= 0.316 \textwidth,clip,trim=3.6cm 0 0 0.8cm]{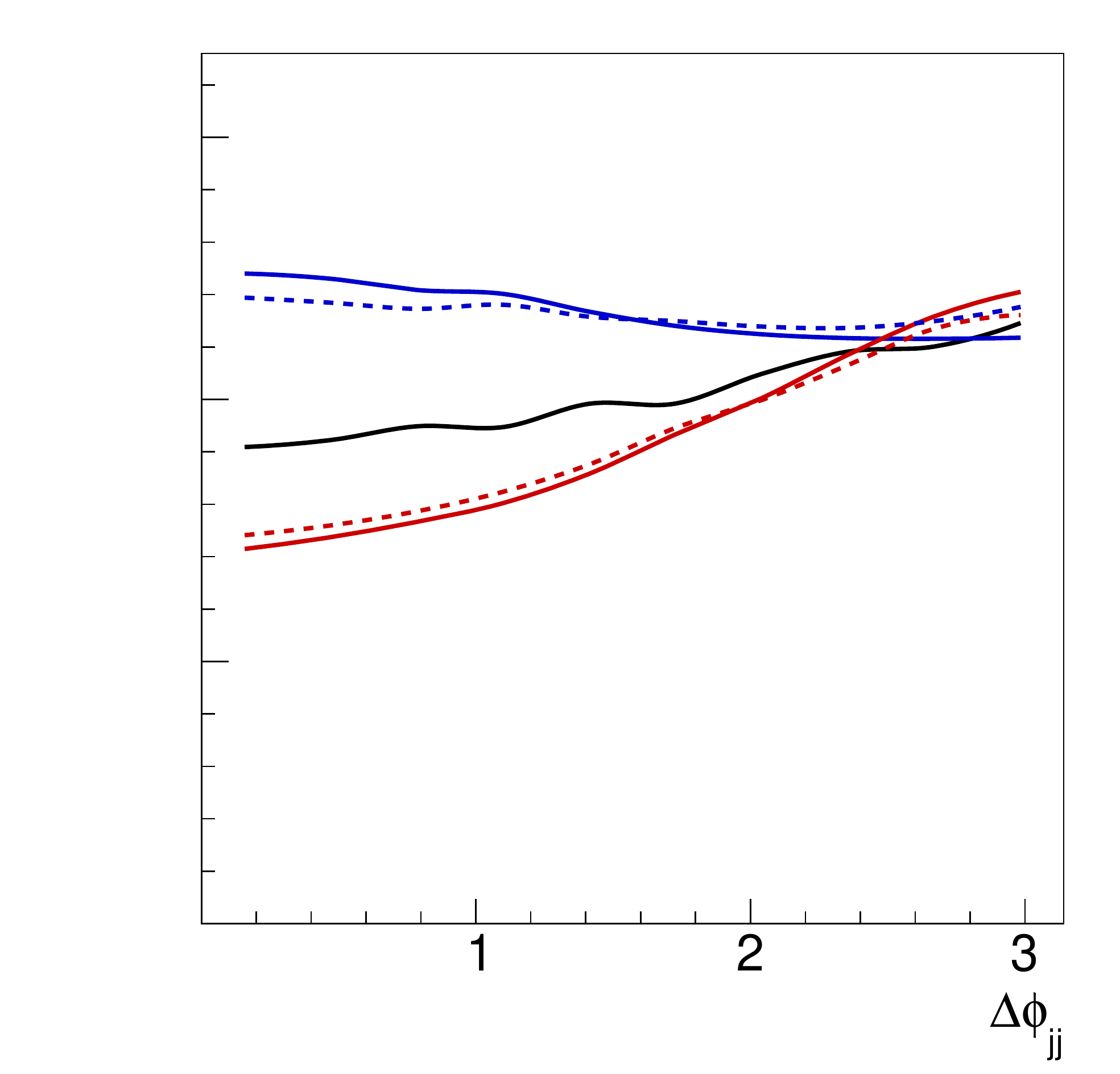} 
\caption{Effects of $\ope_W$ on kinematic distributions and the
  corresponding results from our simple model. For the first we
  separate the first terms (left panels) and second term (middle
  panels) as defined in Eq.~\eqref{eq:OW_HWW}. The right panels
  show the effect of the full operators.}
\label{fig:Dim6_Comparison}
\end{figure}

We can compare the effects of both approaches on our leading kinematic distributions.
Let us first separately consider the modification of
the $HWW$ vertex by the first term and
second term of Eq.~\eqref{eq:OW_HWW}. In both cases we choose
$c_{\phi,2}/\Lambda^2 = \pm 10/\tev^2$, which is slightly outside current
exclusion limits~\cite{Dim6Limits}, in order to make the effect of this operator more visible.
In these two setups we compute cross sections as well as $p_T$ and
$\Delta \phi_{jj}$ distributions.
The results are shown in
Figure~\ref{fig:Dim6_Comparison} as the red and blue solid lines.  

Considering only the first term of Eq.~\eqref{eq:OW_HWW}, i.\,e.\ shifting
the Standard Model coupling strength, we find that the cross section
noticeably increases with positive $c_W$ and decreases with negative
$c_W$ in line with our expectation from
Eq.~\eqref{eq:Dim6EquivalentParam2}.  We can now compare to our simple
model. The dashed lines in the left panels of
Figure~\ref{fig:Dim6_Comparison} correspond to the results of our
simple model with parameters chosen according to
Eq.~\eqref{eq:Dim6EquivalentParam2}.  The agreement is very good, as
one would expect, because the shift in $a_{L,T}$ is
momentum-independent for an on-shell Higgs.  The only differences can arise
from the (small) off-shell Higgs contributions that survive the cuts
we have made.

The second term of Eq.~\eqref{eq:OW_HWW} introduces new Lorentz
structures. It not only modifies the rate, but also changes the
kinematic features. In particular, the couplings $a_{L,T}$ that were
constant in our simple model now depend on the momenta.  Positive
values of $c_W$ now lead to a significantly reduced cross section,
reflecting the relative sign between the two dimension-6
contributions.  They also slightly soften the tagging jets and induce
a preference for back-to-back configurations of the two tagging
jets.  Conversely, negative $c_W$ increase the rate, lead to harder
tagging jets, and favor aligned tagging jets.\bigskip

To compare with our simple model we fit a constant $a_L$ ($a_T=1$ as argued above). 
For instance, for $c_W/\Lambda^2 = 10 /\tev^{2}$
we find that a constant longitudinal coupling $a_L^{(2)}=0.89$ corresponding to
$F(p^+,p^-) = - (105 \ \gev)^2$ yields good agreement in the cross
section and the angular correlation between $\ope_W$ and our simple
model. This can be seen in the bottom middle panel of
Figure~\ref{fig:Dim6_Comparison}.
However, the transverse momentum of
the leading tagging jet, shown in the top right panel of
Figure~\ref{fig:Dim6_Comparison}, indicates substantial deviations.
For positive $c_W$ the tagging jets in the dimension-6 approach are
softer while for negative $c_W$ they are harder than in our simple
model. This reflects the fact that a constant value of $F(p^+,p^-)$
entering $a_L^{(2)}$ does not account for the momentum dependence of
$\ope_W$. \bigskip

Ignoring the momentum dependence, the preferred effective value of $F(p^+,p^-)$
is worth a brief discussion. It describes a contribution to the
longitudinal coupling that is anti-proportional to the effect of the first
term in Eq.~\eqref{eq:OW_HWW}.
After combining, the two modifications of the longitudinal coupling nearly
cancel for $c_W / \Lambda^2 = 10/\tev^2$, while
the transverse coupling is unaffected by the second modification:
\begin{alignat}{5}
a^{(1)}_L + a^{(2)}_L
&= 1 + \frac{c_{W}  }{\Lambda^2} \; \left( \frac 1 2 m_H^2 - (105 \ \gev)^2 \right) 
 = 0.97 \; , \notag \\
a^{(1)}_T + a^{(2)}_T 
&= 1 + \frac{c_{W}  }{2 \Lambda^2} m_H^2
= 1.08 \; .
\label{eq:fit}
\end{alignat}
This means that $\ope_W$ primarily affects the transverse Higgs--gauge
interactions, as expected from the equivalence theorem.\bigskip

A final comparison between the full operator~$\ope_W$ and the
corresponding simple model as shown in the right panels of
Figure~\ref{fig:Dim6_Comparison} supports these observations. The large
effects on the cross sections from the individual two terms, which
correspond to a large modification of the longitudinal coupling,
partly cancel. We are left with a modest decrease of the rate for
positive $c_W$ and an increase for negative $c_W$. The
kinematical distributions follow exactly the same pattern as for the
second term only.  The simple model, where the equivalent coupling
$a_L$ is now close to one, predicts rates and $\Delta \phi_{jj}$
distributions in good agreement with the dimension-6 results.  The
discrepancy in the jet $p_T$ distribution reflecting the missing
momentum dependence remains as well.

All in all we find that while not all details can be directly linked
to higher-dimensional operators, our simple model gives a good
description of the qualitative features in terms of the intuitive
gauge boson polarizations.

\section{Conclusions}
\label{sec:conc}

Based on the Goldstone equivalence theorem, one can argue that
deviations from the Standard Model Higgs couplings to transverse and
longitudinal gauge bosons could have a very distinctive physics
origin. The longitudinal gauge bosons are inherently connected to
features of the Higgs sector, whereas those of the transverse ones test
the nature of the original gauge bosons. 

While such tests of the Higgs--gauge sector are theoretically
and experimentally well
established for the high-energy regime, we propose to extend
this approach to the Higgs resonance. 
Experimentally, this strategy has the benefit of significantly higher
rates, so we can try to search for deviations from the Standard Model
in the bulk of the $VV \to VV$ cross section rather than in its tails.

On the theory side, the separation of gauge boson polarizations away
from the high-energy limit requires some care. In particular it suffers 
from the fact that to perform this separation one has to explicitly 
break Lorentz invariance to define polarizations.
Independent of this issue, the observables we investigate can be viewed
as a first step towards testing such physics and can more generally
be useful in probing the structure of the Higgs--gauge sector.
We have demonstrated this in a comparison with a fully
Lorentz-invariant model based on effective field theory.
In particular, we have shown that
independent couplings of longitudinal and transverse gauge bosons
to the Higgs can be generated in a model described by dimension-six
electroweak operators.

We complement the information from the total rate at the Higgs resonance
with the kinematics of the tagging jets. Both their transverse momenta
and their angular correlation provide valuable information on the structure
of the $HVV$ coupling, parametrized in terms of the longitudinal and
transverse gauge boson polarizations. The constraints from the
kinematical distributions are orthogonal to the constraints from the
cross section.
When it comes to numbers, systematic effects become relevant.
In this first study we have limited ourselves to a parton-level analysis of
the dominant signal and background processes, omitting
theoretical and systematic uncertainties. Combining the cross section
and tagging jet observables, longitudinal and transverse
gauge boson couplings to the Higgs can be individually probed at the
$\ord(20\%)$ level using $300~\ifb$ of integrated luminosity
at 13~TeV.


\end{fmffile}

\end{document}